\documentclass{ws-rv9x6}
\usepackage{subfigure}   
\usepackage{ws-rv-thm}   
\usepackage{ws-rv-van}   
\makeindex

\begin{document}

  \def\apj{ApJ}
  \def\pasp{PASP}
  \def\pasj{PASJ}
  \def\araa{ARA\&A}
  \def\aap{A\&A}
  \def\apjl{ApJL}
  \def\apjs{ApJS}
  \def\mnras{MNRAS}
  \def\aj{AJ}
  \def\nat{Nature}
  \def\grl{GRL}
  \def\icarus{Icarus}
  \def\baas{BAAS}
  \def\ssr{SSR}
  \def\prl{PRL}
  \def\nar{NAR}

\chapter{The Compositional Dimension of Planet Formation}

\author[D. Turrini]{Diego Turrini\footnote{Email: {diego.turrini@inaf.it}}}

\address{INAF - Osservatorio Astrofisico di Torino.}

\begin{abstract}
The great diversity of the thousands of planets known to date is proof of the multitude of ways in which formation and evolution processes can shape the life of planetary systems. Multiple formation and evolution paths, however, can result in the same planetary architecture. Because of this,  unveiling the individual histories of planetary systems and their planets can prove a challenging task. The chemical composition of planets provides us with a guiding light for navigate this challenge, but to understand the information it carries we need to properly link it to the chemical composition and characteristics of the environments in which the planets formed. To achieve this goal it is necessary to combine the information and perspectives provided by a growing number of different fields of study, spanning the whole lifecycle of stars and their planetary systems. The aim of this chapter is to provide the unifying perspective needed to understand and connect such diverse information, and illustrate the process through which we can decode the message contained into the composition of planetary bodies.
\end{abstract}


\body


\section{Introduction}\label{sec:introduction}

After centuries where the Solar System was the only planetary system known to humanity and the template on which we shaped our understanding of planet formation, the past quarter of a century saw our view of planets in our galaxy completely revolutionized. A particularly unexpected finding is that the architectures of the thousands of exoplanetary systems known to date span over more than six orders of magnitude in distance from their host stars, going from planets closers to their stars than Mercury is to the Sun and whose equilibrium temperatures exceed a thousand K degrees, to planets embedded in the cold outer regions of their native protoplanetary disks at hundreds of astronomical units.

The spatial extension of the planet-forming region hinted by this great diversity of orbital architectures goes well beyond what was suggested by the architecture of the Solar System alone, and implies an also great diversity of formation environments in the natal protoplanetary disks. This, in turn, is expected to result in a large variety of compositional natures of the formed planets, as already hinted by the first exoplanetary population studies (see e.g. \cite{thorngren2016,tsiaras2018,pinhas2019,welbanks2019}). A growing body of literature is investigating what kind of constraints can such varied  compositions of the planetary bodies provide on their formation environments and their interplay with the planet formation process (e.g. \cite{turrini2015,madhusudhan2016,turrini2018,madhusudhan2019,oberg2020,turrini2021b} and references therein, and  \cite{boothr2019,oberg2019,bosman2019,cridland2019,turrini2021a,schneider2021a,schneider2021b}).

Understanding the connection between planet formation and planetary composition requires an interdisciplinary point of view encompassing an expanding number of disciplines: from the characterization of stars to the study of the circumstellar disks and the interstellar medium from which they were born, and from the investigation of meteorites and comets in the Solar System to the study of extrasolar materials contaminating the atmospheres of young forming stars as well as of white dwarfs at the end of their evolution. Since discussing each of these fields of study in detail is impossible within the scope of this chapter, the referenced bibliography includes recent reviews on each of these topics for  in-depth discussions.

While this chapter aims to provide an updated view of how the planet formation process and the environments in protoplanetary disks shape the composition of planets, the rapid pace at which the underlying fields of study are expanding continuosly brings new results in the picture. As such, the main goal of this chapter is to provide the unifying view linking the different pieces of information each of these fields of study provide and to illustrate the process through which they can be connected to investigate the compositional dimension of planet formation. Before starting our journey, the following Sects. \ref{sec:planet_formation} and \ref{sec:astrochemistry} will introduce most of the basic concepts that will accompany us throughout the rest of the chapter.


\subsection{Planet formation: basic concepts}\label{sec:planet_formation}

\begin{figure}[ht]
\centering
\includegraphics[width=\textwidth]{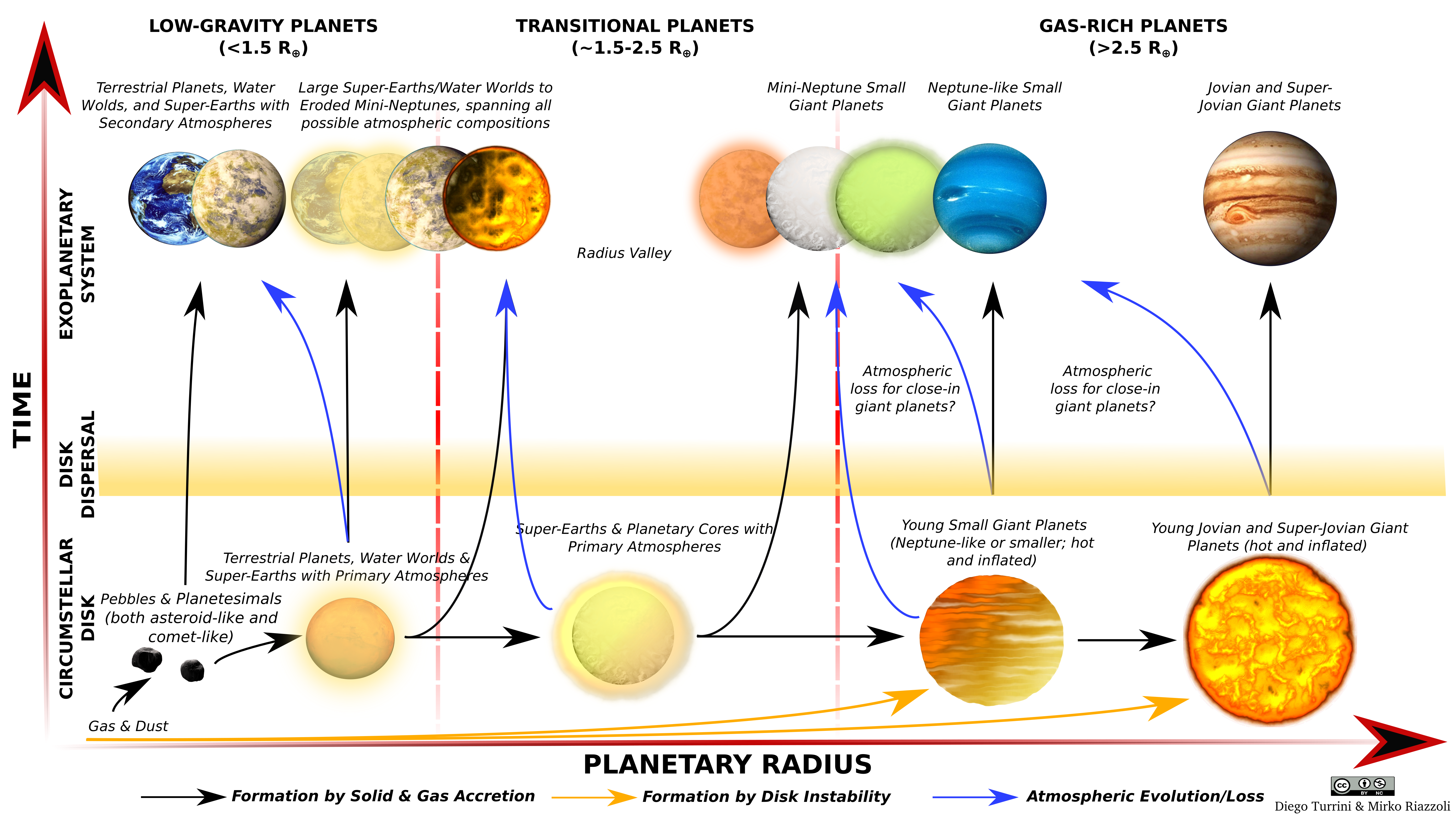}
\caption{Schematic representation of the formation paths that, starting from the gas and dust in circumstellar disks (the bottom left corner of the diagram), create the different kinds of planets currently observed. Orange and black arrows indicate the paths linked to the formation process, i.e. disk instability and solid/core accretion plus gas capture respectively, while blue arrows indicate the paths shaped by atmospheric evolution (e.g. atmospheric escape, atmospheric erosion, outgassing). Planets are divided into three broad categories: high-density planets (mainly
composed by Si, Mg, Fe, C, O), gas-rich planets (for which H and He represent a significant fraction of their mass) and transitional planets (encompassing the transition between the largest high-density planets and the smallest gas-rich planets). Figure updated from\,\cite{turrini2018}}\label{fig:planet_formation}
\end{figure}

The planet and star formation processes are strongly connected, as planets are born in the circumstellar disks surrounding young forming stars. The circumstellar disks share the same bulk elemental composition of their central stars, as they are both composed by the gas and dust inherited from the parent molecular cloud. Planets are born from the dust and gas of circumstellar disks through two main pathways, one involving the gravitational collapse of the disk gas and the other involving the growth and conversion of the dust into larger and larger objects.

The first pathway, indicated by the orange lines in Fig. \ref{fig:planet_formation}, involves the collapse of parcels of disk gas under their own self-gravity: this process requires cold and massive disks to be triggered, is more effective in forming planets at large distances ($\gtrsim$ 50 au) from the central star where the temperatures are lower and results in massive planets with masses comparable or greater than that of Jupiter \cite{dangelo2010,helled2014,turrini2018}. The second pathway, indicated by the black lines in Fig. \ref{fig:planet_formation}, involves the gradual growth of dust into larger and larger bodies\cite{dangelo2010,helled2014,turrini2018}. 

Collisions among dust grains will result in their sticking and growth due to electromagnetic forces producing pebbles up to about cm size, after which collisions become erosive and can result either in mass conservation or mass loss\cite{johansen2014,johansen2017}. Dust and pebbles of different sizes will be affected at different levels by their interaction with the disk gas\cite{weidenschilling1977} (see Sect. \ref{sec:dust_and_planetesimals} for further details), resulting in their differential motion and favouring their concentration and clustering. Clusters of dust and pebbles can become gravitationally bound and collapse under their own self-gravity, forming bodies with sizes ranging from tens to thousands of km in size called planetesimals\cite{chambers2010,johansen2014,johansen2017} on timescales of the order of 10$^5$-10$^6$ years (see \cite{scott2007,nittler2016,wadhwa2020} for the meteoritic constraints on the timescale of this process). 

During the lifetime of circumstellar disks, planetesimals will then grow due to both mutual low velocity collisions and the accretion of pebbles to give rise to larger bodies with masses of the order of $\sim$0.01-0.1 Earth masses, the planetary embryos\cite{chambers2010b,johansen2019} (see \cite{brasser2013,lammer2021} for the constraints from the Solar System on the timescale of this process). The planetary embryos will also grow due to the accretion of pebbles (while still embedded in the circumstellar disk) and mutual collisions (predominantly after the disk dispersal) until they give rise to larger solid bodies with masses spanning between those of the terrestrial planets in the Solar System and those of the super-Earths planets observed around other stars\cite{chambers2010b,turrini2018}.

The timescale on which this process completes can vary between a few 10$^6$ to several 10$^7$ years (see \cite{chambers2010b,brasser2013,lammer2021} for the constraints from the Solar System on the timescale of this process) so it can significantly exceed the lifetime of circumstellar disks, which is of the order of a few 10$^6$ years \cite{meyer2009,fedele2010}. Planetary bodies succeeding in growing to masses of the order of $\sim$10 Earth masses before the dispersal of their protoplanetary disk will be able to capture significant masses of gas from the disk in the form of expanded atmospheres\cite{chambers2010b,dangelo2010,helled2014,turrini2018}. Based on their composition being linked to that of the protoplanetary disk, these atmospheres are called primary atmospheres to differentiate them from the secondary atmospheres produced by volcanism and outgassing processes during the geophysical evolution of planets\cite{turrini2018}. 

If the mass of captured gas becomes comparable to that of the planetary body to which it is gravitationally bound, the expanded atmosphere will begin to collapse, triggering a runaway process that will allow the solid planet to grow by one to two orders of magnitude in mass and become the core of a giant planet\cite{dangelo2010,helled2014,turrini2018}, as illustrated by the horizontal branch of the planet formation sequence (black lines) preceeding the disk dispersal in Fig. \ref{fig:planet_formation}. Differently by the formation pathway shaped by disk instability (orange lines in Fig. \ref{fig:planet_formation}), the core instability pathway for forming giant planets favours orbital regions closer to the central star ($\lesssim$ 50 au)\cite{dangelo2010,helled2014,turrini2018} due to their higher densities of solids (see Sect. \ref{sec:protoplanetary_disks}).

While the first pathway will result in massive giant planets, planets forming through the second pathway will not necessarily become giant planets. The growth process from dust to planetary objects can stop at any stage, resulting in a continuum of planetary masses spanning across the whole spectrum sampled by the terrestrial planets in the Solar System and the super-Earths observed around other stars as illustrated in Fig. \ref{fig:planet_formation}. Composition-wise, it is important to emphasize that this process of growth is not confined to the local materials initially surrounding the seeds of the forming planets\cite{mordasini2016,turrini2018}.

\begin{figure}[ht]
\centering
\includegraphics[width=\textwidth]{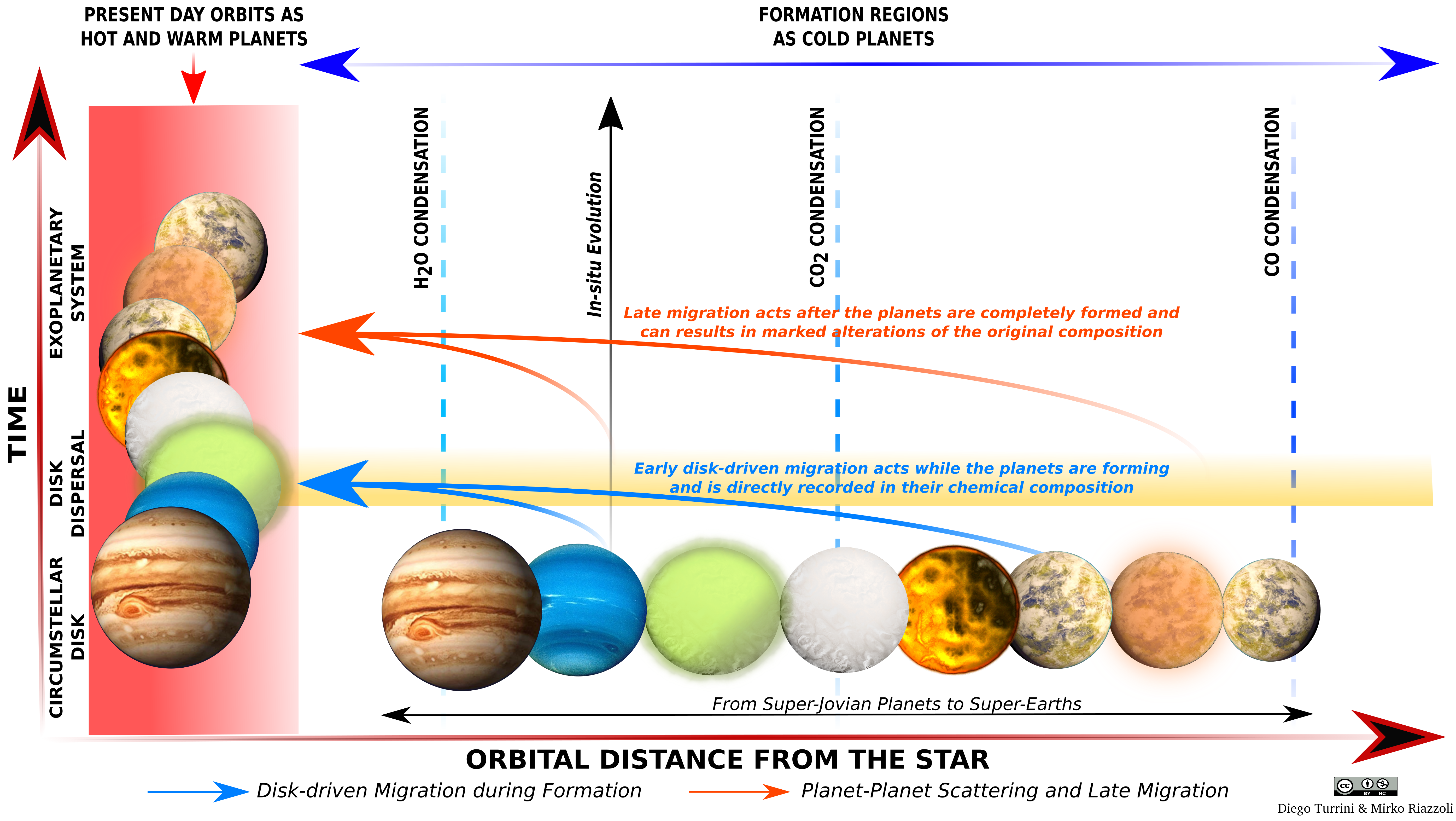}
\caption{Schematic representation of the dynamical pathways that can deliver planets from their formation regions across the extension of the natal protoplanetary disk to the orbital regions closest to the star. While embedded in their native protoplanetary disks, planets can experience early disk-driven migration due to their interactions with the surrounding gas. Dynamical chaos and planet-planet scattering events can later affect the architecture of planetary systems and cause their planets to experience late
migration. Figure updated from\,\cite{turrini2018}}\label{fig:planet_migration}
\end{figure}

As we will discuss in more details in Sect. \ref{sec:dust_and_planetesimals}, dust and pebbles drift radially across the circumstellar disk due to their interaction with the disk gas. Likewise, during their growth the various planetary bodies will interact with the gas in the disk and will migrate radially\cite{chambers2009,nelson2018}, crossing different compositional regions of the disk (see Fig. \ref{fig:planet_migration}). When most of the initial mass of dust in the disk is converted into a limited number of large planetary bodies, their mutual gravitational interactions will cause them to be scattered (either inward or outward) from their original orbit\cite{chambers2009,turrini2015,turrini2018} and will allow for planetary bodies initially distant from each others to collide and compositionally mix (see Fig. \ref{fig:planet_migration} and\,\cite{turrini2015}). 

As a result, the orbital location where we observe a planet can in most cases bear little to no significance for identifying the formation region of the planet and constrain its past dynamical evolution\cite{turrini2015}. The density of planets can provide us insight on the abundances of refractory and volatile elements in their elemental budgets and, as such, put constraints on their formation regions or accretion histories, though this information is often uncertain and degenerate\cite{thorngren2016,turrini2018,adibekyan2021}. In order to further advance our understanding of planet formation, we therefore need to decode the information enclosed in the composition of planetary bodies\cite{turrini2015,madhusudhan2016,turrini2018,madhusudhan2019}. To properly interpret the compositional signatures left by the planet formation process, in turn, we need to characterize the environment in which planets form\cite{turrini2021b}.

\subsection{Chemistry in planet formation: basic concepts}\label{sec:astrochemistry}

The next step before starting the journey into the compositional dimension of planet formation is to make order into the sometime confusing overlap between the nomenclature and classification schemes adopted in the two main disciplines contributing to our understanding of the subject, astrochemistry and cosmochemistry. The guiding principle at the basis of both classification schemes, however, is always  the comparison of the relative volatility of different elements. 

To first order, elements can be divided into two broad categories: refractory and volatile elements. This broad division, adopted in astrochemical studies of protoplanetary disks, is based on the condensation behaviour with temperature of the elements. \textit{Refractory elements} condense at higher temperature than volatile elements and they are in solid form across most of the extension of protoplanetary disks. In the following we will refer to solids mainly composed of refractory elements as refractory materials, grouping both \textit{rocks and metals} in this general category. \textit{Volatile elements} remain in gas form down to lower temperatures so, opposite to refractory elements, they are present as gas across vast regions of protoplanetary disks. Volatile elements condense as \textit{ices}. Throughout the chapter, we will generally follow this very broad classification unless otherwise specified.

The transition between refractory and volatile elements can be placed roughly around 300 K\cite{lodders2003,fegley2010}. As we will see in Sect. \ref{sec:planetary_materials}, the meteoritic record in the Solar System\cite{lodders2003,lodders2010a,palme2014,palme2017} shows us that at this temperature all elements except hydrogen (H), the noble gases and the volatile elements carbon (C), oxygen (O) and nitrogen (N) are in condensed form as rocks and metals. Molecules composed by volatile elements can be differentiated between volatile molecules (e.g. H$_2$O, NH$_3$, CO$_2$) and ultra-volatile molecules (e.g. CH$_4$, CO, N$_2$) with the transition roughly occurring around 30 K\cite{lodders2003,fegley2010}. The noble gases can condense as ices at temperatures lower than 30 K, the condensation of Ne requiring temperatures of about 9 K to occur\cite{fegley2010}. The noble gases Ar, Kr and Xe can also be effectively trapped into H$_2$O ice as \textit{clathrates}\cite{fegley2010} at comparatively higher temperatures if the abundance of water ice is large enough.

As we will further discuss in Sect. \ref{sec:planetary_materials}, the volatile elements O and C actually behave both as refractory and volatile elements. O is distributed in comparable quantities between refractory materials (rocks are mainly composed of Fe, Si and Mg oxides\cite{fegley2010}) and ices (H$_2$O, CO, CO$_2$)\cite{lodders2010a,palme2014,palme2017}. C is present in limited quantities in refractory materials\cite{lodders2010a,palme2014,palme2017} but is the major building block of \textit{organics}, which in turn can be divided into refractory organics and volatile organics\cite{pollack1994,semenov2003,thiabaud2014,bergin2015,mordasini2016,altwegg2019,oberg2020,turrini2021a,turrini2021b}. The third volatile element, N, is that characterized by the highest volatily and only a minor fraction condenses with refractory materials and organics\cite{lodders2003,gibb2004,lodders2010a,palme2014,palme2017,oberg2019,altwegg2019,oberg2020}. 

The categorization discussed until now adopts the perspective of the field of study of astrochemistry. The source of confusion mentioned at the beginning arises from the fact that the second field of study dealing with the chemistry of planetary systems, cosmochemistry, uses of a similar terminology to indicate somewhat different classes of elements. The starting point of cosmochemical studies are the samples of planetary materials, so there is a strong focus on the composition of solids in protoplanetary disks. Astrochemical studies instead have their starting point in the astronomical observations of molecular clouds and protoplanetary disks, therefore putting a larger emphasis on the volatile elements.

Within the cosmochemical classification, the elements astrochemically classified as volatile elements are defined as both highly volatile and atmophile elements\cite{fegley2010,lodders2010a}, since their high volatility makes them the major component of planetary atmospheres. The astrochemical category of refractory elements, instead, is split into four different classes. The cosmochemical class of \textit{refractory elements} groups those elements condensing at temperatures higher than 1360 K (further subdivisions based on the condensation temperature and the chemical behaviour exist and are discussed in \cite{lodders2003,fegley2010} but are beyond the scope of this chapter), with Ca, Al and Ti being the most abundant refractory elements.

The next group includes elements condensing between 1360 K and 1290 K and contains Fe, Si and Mg: these are among the most abundant cosmic elements and, together with O, the major components of rocks, which is why this group is called \textit{major} or \textit{common elements}\cite{lodders2003,fegley2010}. In order of decreasing condensation temperatures we then encounter the moderately volatile elements, which condense at temperatures above 700 K and include P, F, Cl, Na and K, and the volatile elements, which include S and condense at temperatures between 700 and 300 K\cite{lodders2003,fegley2010}. The condensation temperature of S as FeS (troilite) at about 700 K at is what marks the transition between moderately volatile and volatile elements\cite{lodders2003,fegley2010}. 

Before proceeding, it is important to point out that condensation is a gradual process where the balance between the condensed and gaseous phases varies over a range of temperature values (at constant pressure)\cite{fegley2010,madhusudhan2016}. This, in turn, implies that the condensation fronts (also called \textit{snow lines}) are transition regions that can extend over multiple au rather than sharp boundaries. When only one condensation temperature is reported for a element or molecule, its value implicitly refers to the condensation of 50$\%$ of its total mass\cite{lodders2003,fegley2010,lodders2010a}. Finally, it is also important to point out that the concept of snowline is not always as well defined as one would expect: in chemically active protoplanetary disks, molecules can be characterized by wavy condensation patterns and actually have multiple snow lines\cite{eistrup2016,eistrup2018}.

\section{The initial chemical budget of planet formation}

The first step to understand planetary composition and the information it carries on planet formation lies in characterizing the initial elemental budget contained in the protoplanetary disks from which planets are born. In the case of young planets still embedded in the native protoplanetary disks, it is in principle possible to directly access this record by probing the composition of the dust and gas components of the disks themselves. 

The original protoplanetary disks of the more than 3500 exoplanetary systems currently known as well as that of the Solar System, however, have long dispersed. Nevertheless, records of their composition are still provided by their stars and by the composition of their planetary bodies. Combining the information on the stellar composition and that of the planetary bodies can allow to reconstruct the initial chemical budget and structure of the original protoplanetary disks.

\subsection{Protoplanetary disks}\label{sec:protoplanetary_disks}

Protoplanetary disks are flat, extended structures surrounding young forming stars and composed by the gas and dust inherited from the parent molecular cloud. Supported by their rotation againt the stellar gravity and irradiated by their central star, protoplanetary disks are characterized by pressure, density and temperature gradients along the radial direction from the host star and the vertical direction from their midplanes. These gradients  affect the local thermochemical equilibrium and result in the existence of different compositional regions across their extensions. Here we will focus on describing the physical and thermal structures of disks, to pave the road for discussing their compositional structure in Sect. \ref{sec:compositional_structure}.

In their vertical direction, protoplanetary disks are characterized by three main chemical layers\cite{turrini2021b}: the photon-dominated layer, the warm molecular layer, and the cold midplane. The photon-dominated layer is the outermost vertical layer: here the disk gas interacts with the stellar and interstellar radiation, is mostly in atomic form due to the photo-dissociation of molecules and its chemistry is regulated by photo-processes\cite{turrini2021b}. The warm molecular layer is the intermediate vertical layer: it is shielded from the photo-dissociative radiation and its gas is rich in molecular species produced through ion-neutral and neutral-neutral reactions\cite{turrini2021b}.

The cold midplane is the layer most directly involved in planet formation, it is the innermost layer and as such is shielded from the direct stellar radiation by the other layers\cite{turrini2021b}. Due to the aerodynamical drag exerted by the gas, the dust originally mixed with the gas throughout the whole vertical extension of the disk settles in this layer, resulting in the highest density of solids\cite{testi2014}. As a consequence, the chemistry in the midplane is mainly controlled by the condensation of the chemical species on the surface of the dust grains and by grain-surface reactions\cite{oberg2020,turrini2021b}. In the following we will focus our attention on the disk midplane, though it is important to bear in mind that the warm molecular layer is expected to play a role in the accretion of the gaseous envelope of giant planets\cite{oberg2020}.

On the disk midplane, with the exception of the innermost region (generally of the order of 1 au\cite{ida2016}), the temperature gradient in the radial direction depends on the distance from the host star as 
\begin{equation}
T_{m} = T_{0} \left( r/r_{0} \right)^{-\beta}\label{eqn:temperature_profile}
\end{equation} 
where T$_{0}$ is the normalization temperature at the reference distance r$_{0}$, and the exponent $\beta$ is generally found to be close to 0.5. Generally, in studies of the Solar System the reference distance is 1 au while in observational studies of circumstellar disks the reference distance is much larger and depends on the spatial resolution of the observations: e.g. \cite{isella2016} adopt a reference distance of 100 au for the circumstellar disk surrounding the star HD\,163296.

A classical parametrization of the temperature profile of the original circumsolar disk, also called the solar nebula, adopts a normalization temperature T$_0$=280 K at r$_0$=1 au with $\beta$=0.5 based on the observation of the distribution of refractory and volatile elements in the present Solar System\cite{hayashi1981}. Such parametrization, still widely used in present-day studies, adopts a midplane temperature profile warmer than those observationally estimated for circumstellar disks. As a comparison, the circumstellar disks around HD\,163296, AS 209 and TW Hya are characterized by normalization temperatures at 1 au of 240 K, 150 K and 130 K respectively\cite{isella2016,fedele2016,huang2018,favre2019}. The $\beta$ values of these disks are instead close to that adopted for the solar nebula, being 0.5-0.6 for HD\,163296 and AS 209, and 0.47 for TW Hya\cite{isella2016,fedele2016,huang2018,favre2019}. 

\begin{table}[t]
\centering
\tbl{Snowlines of volatile molecules.}
{\begin{tabular}{@{}cccc@{}} 
\toprule
Molecule & 50\% T$_C$ (K) & 50\% R$_C$ (au) & Reference \\
\colrule
H$_2$O   & 140   &  3        &  \cite{martin-domenech2014} \\
CH$_3$OH & 102.5 &  6.5      &  \cite{martin-domenech2014}\\
NH$_3$      &  80   &  9        &  \cite{martin-domenech2014}\\
CO$_2$      &  65   &  13.7     &  \cite{martin-domenech2014}\\
CO       &  30   &  64     &  \cite{fayolle2016,oberg2019}\\
N$_2$       &  26 &  85      &  \cite{fayolle2016,oberg2019}\\
\botrule 
\end{tabular}}
\begin{tabnote}
Snow lines of the most abundant volatile molecules for the disk temperature profile of HD\,163296, where the condensation distance R$_C$ is determined by the orbital distance where the disk temperature matches the condensation temperature T$_C$ at which 50\% of the volatile molecule condenses as ice.
\end{tabnote}
\label{tab:snowlines}
\end{table}

Adopting as case study the midplane thermal profile of HD\,163296 and the condensation temperatures of the abundant volatile molecules listed in Table \ref{tab:snowlines}, we can have a first dive into the compositional structure in the disk midplane. With the reference temperature T$_{0}$=240 K at 1 au, all elements belonging to the astrochemical refractory family will be condensed in solid form already at about 0.6 au (300 K, see Section \ref{sec:astrochemistry}). Moving away from the star we first encounter the snow line of H$_2$O, the most refractory/least volatile molecule formed by the volatile elements, at about 3 au.

The next snowline we encounter is that of the volatile organic molecule CH$_3$OH at 6.5 au, followed by those of the ices NH$_3$ at 9 au and CO$_2$ at 13.7 au (see Table \ref{tab:snowlines}), i.e. from two to four times more distant from the star than the snow line of H$_2$O. To encounter the snow lines of the highly volatile molecules CO and N$_2$ we need to go five to six times more distant than the snow line of CO$_2$, reaching 64 and 85 au respectively (see Table \ref{tab:snowlines}). The most abundant elements H and He, as well as the noble gas Ne (see Sects. \ref{sec:astrochemistry} and \ref{sec:host_stars}), never condense in circumstellar disks and always remain in gas form.

As this illustrative example showcases, the different volatility of the elements and of their carriers (rocks, organics, ices) results in compositional gradients across the midplane of circumstellar disks: as we will see in Sect. \ref{sec:compositional_structure}, the farther we move from the star, the colder the local regions of the disk are, and the more volatile elements and molecules will condense adding to the mass fraction of solids. The colder the circumstellar disk, the closer the snow lines will be to the star: for a temperature profile similar to that of AS 209 the snow lines of H$_2$O and N$_2$ would be at $\approx$1.2 and 33 au respectively, i.e. 2.5 times closer to the host star.

To understand how abundant the different solid components are across the circumstellar disk, the information provided by the temperature profile needs to be combined with the one on the mass distribution provided by gas surface density profile. Based on observational data\cite{andrews2010,isella2016,miotello2018}, the surface density profile of circumstellar disks can be expressed as:
\begin{equation}
\Sigma_{gas}(r) = \Sigma_{c} \left( r/r_{c})\right)^{-\gamma} exp \left[ -\left( r/r_{c} \right)^{2-\gamma}\right]\label{eqn:gas_density_tapering}
\end{equation}
where r$_c$ is the characteristic radius of the circumstellar disk, $\Sigma_{c}$ the surface density at the characteristic radius and the exponent $\gamma$ has been observational constrained in the range 0.8-1\cite{andrews2010,isella2016,miotello2018}.

\begin{figure}[t]
\centering
\includegraphics[width=\textwidth]{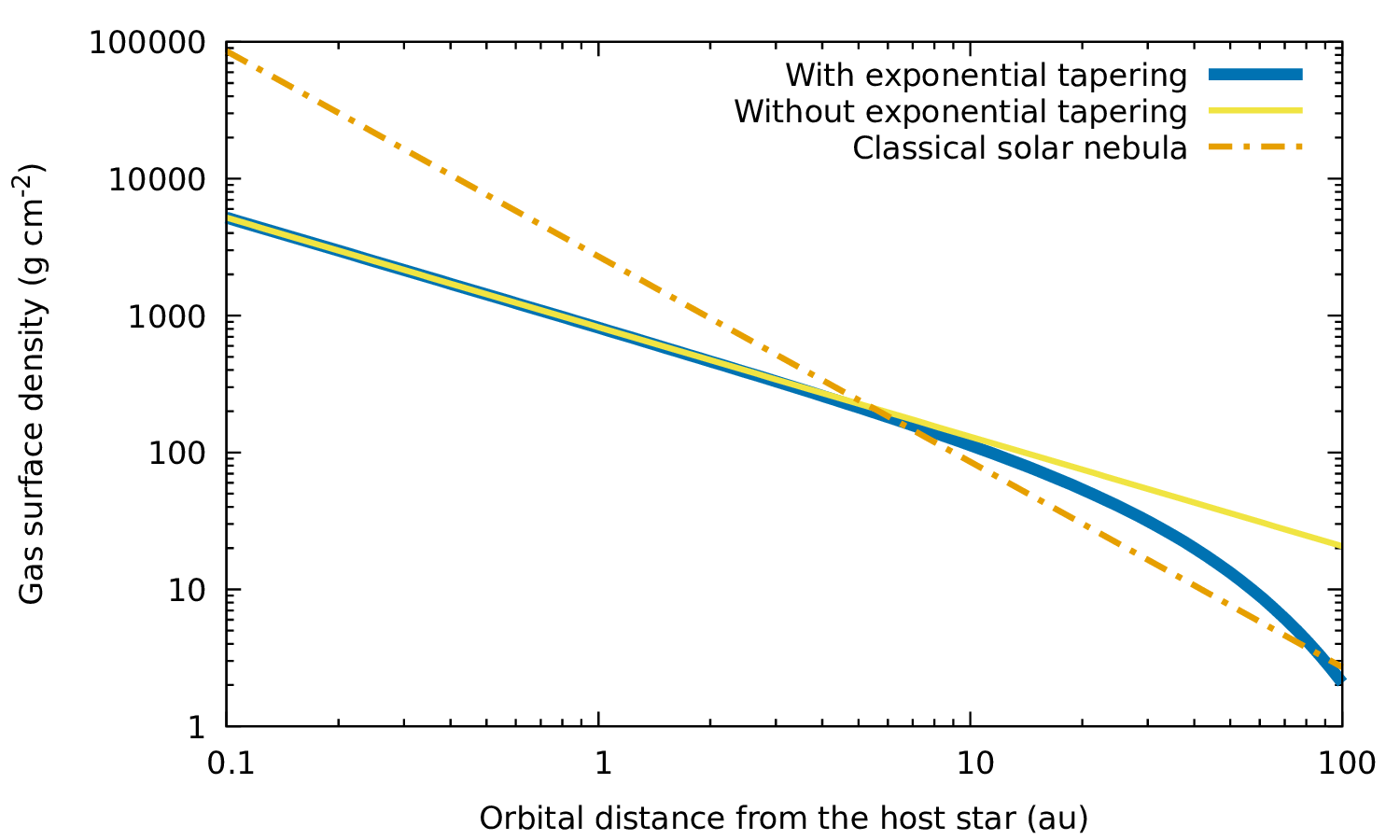}
\caption{Comparison between the gas surface density profiles of circumstellar disks with the same total mass of 0.05 M$_\odot$ using different parametrizations: the blue thick solid curve represents the surface density profile obtained from Eq. \ref{eqn:gas_density_tapering} with $\gamma$=0.8, the yellow thin solid line the surface density profile obtained from Eq. \ref{eqn:gas_density_notapering} with $\gamma$=0.8, and the orange dot-dashed line is the classical solar nebula profile from Eq. \ref{eqn:gas_density_notapering} and $\gamma$=1.5\cite{coradini1981,hayashi1981}.}\label{fig:disk_density_profiles}
\end{figure}

Also in the case of the gas surface density profile, a number of present-day studies still adopt the classical parametrization derived in the case of the solar nebula\cite{coradini1981,hayashi1981}:
\begin{equation}
\Sigma = \Sigma_{c} \left( r/r_{c})\right)^{-\gamma}\label{eqn:gas_density_notapering}
\end{equation}
where r$_{c}$ is generally 1 au, $\Sigma_{c}$ is about 2-3$\times10^{3}$ g\,cm$^{-2}$\cite{coradini1981,hayashi1981,chambers2010} and, more importantly, $\gamma$=1.5\cite{coradini1981,hayashi1981,chambers2010}. As illustrated by Fig. \ref{fig:disk_density_profiles}, the classical solar nebula profile is much steeper than those derived from astronomical observations of circumstellar disks. As a consequence of the more rapid fall of its gas surface density, the classical solar nebula profile is characterized by a higher concentration of the disk mass in the regions closer to the star.

The dust density profile, i.e. the initial distribution of solids, is generally obtained from the gas surface density profile by adopting a dust-to-gas ratio $\xi(r)$ across the extension of the protoplanetary disk:
\begin{equation}
\Sigma_{dust}(r) = \xi(r) \cdot \Sigma_{gas}(r)\label{eqn:dust_density}
\end{equation}
where a constant value $\xi(r)=0.01$ derived from observation of the interstellar medium\cite{bohlin1978} is generally assumed\cite{turrini2021b}. As we will see in Sect, \ref{sec:compositional_structure}, even at the beginning of the life of protoplanetary disks $\xi(r)$ is not constant throughout the disk but dependes on the local mass fraction of condensed solids. Nevertheless, this approximation is reasonably accurate to estimate the global dust/solid mass in protoplanetary disks (see also Sect. \ref{sec:host_stars}).
 
Eq. \ref{eqn:dust_density} highlights how there is a direct proportionality between the gas and dust surface density profiles. Since the classical solar nebula is characterized by a higher concentration of gas in the regions closer to the star (see Fig. \ref{fig:disk_density_profiles}) and since these regions are also the hotter ones (see Eq.   \ref{eqn:temperature_profile}) where fewer elements are condensed, the classical solar nebula profile is characterized by a larger abundance of refractory dust and a lower abundance of ice-rich dust compared to observationally-constrained circumstellar disks. This translates in the classical solar nebula being a comparatively less favourable environment for the formation of volatile-rich planetary bodies with respect to more realistics circumstellar disks.

\subsection{The host stars and their composition}\label{sec:host_stars}

The composition of stars is characterized by observationally measuring the abundances of the elements in their photospheres. Stellar abundances are expressed in what is known as the astrochemical abundance scale\cite{asplund2009,lodders2010a}, where the abundances of the different elements are expressed in logarithmic scale in comparison to hydrogen, the most cosmically abundant element. The astrophysical abundance scale is a comparative scale that adopts as reference the abundance [H] of $10^{12}$ H atoms, meaning that the abundance of H is by definition 12 dex ($log_{10}[H]$). The relative concentration [X/H] of any given elements with respect to H is then expressed as [X/H]=$10^{X-H}$: in the case of oxygen, using the values from Table \ref{tab:protosolar_abundances}, [O/H]=$10^{8.73-12}$=$10^{-3.27}$=5.37$\times$10$^{-4}$.

A subtle yet important point to make is that, as mentioned above, the stellar abundances we can observationally characterized are the photospheric abundances of stars. These abundances are related to, but are not the same as, the abundances of the circumstellar disks that surrounded the stars while they were forming. During the life of a star elements heavier than hydrogen are affected by sinking and over time they get slowly depleted from the stellar photosphere: in the case of the Sun, this effect is estimated to be of the order of 10\%, i.e. 0.04-0.05 dex \cite{asplund2009,lodders2010a}. Correcting for this effect allows to derive the abundances of circumstellar disks from those measured in stellar photospheres.

\begin{table}[t]
\centering
\tbl{Protosolar abundances of the elements}
{\begin{tabular}{@{}cccc@{}} 
\toprule
Element & Protosolar Abundance &  Mass Fraction & [X/H]\\
\colrule
H & 12.00 & 7.15E-01 & 1.00E+00 \\
He & 10.98 & 2.71E-01 & 9.55E-02 \\
O & 8.73 & 6.09E-03 & 5.37E-04 \\
C & 8.47 & 2.51E-03 & 2.95E-04 \\
Ne & 7.97 & 1.34E-03 & 9.33E-05 \\
N & 7.87 & 7.36E-04 & 7.41E-05 \\
Mg & 7.63 & 7.35E-04 & 4.27E-05 \\
Si & 7.55 & 7.07E-04 & 3.55E-05 \\
Fe & 7.51 & 1.28E-03 & 3.24E-05 \\
S & 7.16 & 3.29E-04 & 1.45E-05 \\
Al & 6.47 & 5.65E-05 & 2.95E-06 \\
Ar & 6.44 & 7.80E-05 & 2.75E-06 \\
Ca & 6.36 & 6.51E-05 & 2.29E-06 \\
Na & 6.25 & 2.90E-05 & 1.78E-06 \\
Ni & 6.24 & 7.23E-05 & 1.74E-06 \\
Cr & 5.66 & 1.69E-05 & 4.57E-07 \\
Cl & 5.54 & 8.72E-06 & 3.47E-07 \\
Mn & 5.46 & 1.12E-05 & 2.88E-07 \\
P & 5.45 & 6.19E-06 & 2.82E-07 \\
K & 5.08 & 3.33E-06 & 1.20E-07 \\
\botrule 
\end{tabular}}
\begin{tabnote}
Protosolar abundances of the twenty most abundant elements, their mass fractions in the protosolar mixture and their concentrations with respect to H. Values from\,\cite{asplund2009,scott2015a,scott2015b}
\end{tabnote}
\label{tab:protosolar_abundances}
\end{table}

Table \ref{tab:protosolar_abundances} shows the protosolar abundances and the associated [X/H] relative abundances of the twenty most abundant elements, obtained through the previous procedure from recent estimates of the solar photospheric abundances \cite{asplund2009,scott2015a,scott2015b}. Table \ref{tab:protosolar_abundances} also shows the mass fraction of each element in the protosolar mixture. Mass fractions are obtained by multiplying the abundance of each element by its atomic weight and dividing the result by the sum of the same product over all elements in the stellar mixture.

The information provided by the mass fractions is used to characterize the stellar composition by means of the three parameters X, Y and Z. X is the mass fraction of hydrogen, Y is the mass fraction of helium, and Z is the sum of the mass fractions of all remaining elements, also called the \textit{metallicity}\cite{lodders2010a,asplund2009}. From Table \ref{tab:protosolar_abundances} we can see that in the protosolar mixture H accounts for 71.5\% (X=0.715) of the total mass, He accounts for 27.1\% (Y=0.271) and the metallicity, i.e. all other elements, accounts for only 1.4\% (Z=0.014). For comparison, in the solar photospheric mixture X=0.738, Y=0.249 and Z=0.013\cite{asplund2009,scott2015a,scott2015b}.

The value of the metallicity Z is an important parameter for planet formation as it represents the upper limit to the solid mass fraction that can be reached in circumstellar disks if all elements except for H and He condense. At the beginning of the life of protoplanetary disks the solid mass fraction Z$_{solids}(r)$ matches the dust-to-gas $\xi(r)$, so in the following we will use the two terms interchangeably. As different elements condense at different temperatures the solid mass fraction Z$_{solids}(r)$ will always be lower than the metallicity Z (see Sects. \ref{sec:astrochemistry} and \ref{sec:protoplanetary_disks}). This will affect the amount of solid material available to form terrestrial planets and the cores of giant planets as discussed in Sects. \ref{sec:planet_formation} and \ref{sec:compositional_structure}.

The case of neon is illustrative of this point: Ne represents about 10\% of the metallicity Z in the protosolar mixture (see Table \ref{tab:protosolar_abundances}) and condenses as ice at about 9 K. It is doubtful that the temperature in circumstellar disks can reach such low values, as even in the case of the cold disk AS 209 discussed in Sect. \ref{sec:protoplanetary_disks} such temperature would be reached only at about 280 au. As a consequence, the dust-to-gas ratio in circumstellar disks matching the protosolar composition will reach at most 1.28\% or 0.0128. Note that this argument applies to the initial dust-to-gas ratio of circumstellar disks (see Sect. \ref{sec:protoplanetary_disks}): as introduced in Sect. \ref{sec:planet_formation} and further discussed in Sect. \ref{sec:compositional_structure}, during the lifetime of disks the interplay between dust drift and dust growth will cause the differential migration between gas and dust and between dust grains of different sizes, resulting in locally higher or lower dust-to-gas ratios (see Sect. \ref{sec:compositional_structure} and\, \cite{isella2016} for an illustrative example).

As we will see in Sect. \ref{sec:compositional_structure}, the information supplied by the relative abundances [X/H] and the mass fractions of the different elements can be combined with that on the gas surface density and temperature profiles to build the gas and dust abundance profiles in circumstellar disks. Before proceeding, however, it is important to emphasize that, while the general considerations discussed in this section are valid for all stars and circumstellar disks, the protosolar composition cannot be used to universally characterize the now-dispersed circumstellar disks around other stars.

Depending on their formation environment and their formation age with respect to the formation of the Milky Way, stars will possess different overall metallicities as well as different abundance ratios between their elements\cite{turrini2021b}. As a result of the galactic chemical evolution, stars that formed earlier than the Sun will be generally characterized by lower metallicity values than the Sun, while younger stars will possess higher metallicity values. The spread in metallicity values for the currently known planet-hosting stars is of the order of 0.5 dex or a factor of 3.5 (see \cite{turrini2021b} and references therein).

As the metallicity, also the ratios between their different elements, e.g. the carbon-to-oxygen ratio C/O and the magnesium-to-silicon ratio Mg/Si, will vary from star to star: expected variations are of the order of 0.3-0.4 dex or between a factor of 2 to 2.5 (see \cite{turrini2021b} and references therein). These differences in the initial elemental budget of their circumstellar disks will translate in different chemical evolution histories of the disks and in different elemental partitioning  between gas and solids across their extension. This, in turn, will translate into different compositions of the formed planets.

\subsection{Meteorites, comets and extrasolar materials}\label{sec:planetary_materials}

\begin{table}[t]
\centering
\tbl{Meteoritic abundances of the elements}
{\begin{tabular}{@{}ccccc@{}} 
\toprule
Element & Meteoritic &  Mass & [X/H]* & Meteoritic/Solar\\
        & Abundance & Fraction & & Abundance Ratio\\
\colrule
H & 8.26 & 1.94E-02 & 1.82E-04 & 1.82E-04 \\
He & 1.33 & 9.06E-09 & 2.14E-11 & 2.24E-10 \\
O & 8.44 & 4.66E-01 & 2.75E-04 & 5.12E-01 \\
C & 7.43 & 3.42E-02 & 2.69E-05 & 9.12E-02 \\
Ne & -1.08 & 1.78E-10 & 8.32E-14 & 8.92E-10 \\
N & 6.30 & 2.96E-03 & 2.00E-06 & 2.70E-02 \\
Mg & 7.57 & 9.56E-02 & 3.72E-05 & 8.71E-01 \\
Si & 7.55 & 1.05E-01 & 3.55E-05 & 1.00E+00 \\
Fe & 7.49 & 1.83E-01 & 3.09E-05 & 9.54E-01 \\
S & 7.19 & 5.26E-02 & 1.55E-05 & 1.07E+00 \\
Al & 6.47 & 8.43E-03 & 2.95E-06 & 1.00E+00 \\
Ar & -0.46 & 1.47E-09 & 3.47E-13 & 1.26E-07 \\
Ca & 6.33 & 9.07E-03 & 2.14E-06 & 9.34E-01 \\
Na & 6.31 & 4.97E-03 & 2.04E-06 & 1.15E+00 \\
Ni & 6.24 & 1.08E-02 & 1.74E-06 & 1.00E+00 \\
Cr & 5.68 & 2.63E-03 & 4.79E-07 & 1.05E+00 \\
Cl & 5.27 & 6.99E-04 & 1.86E-07 & 5.36E-01 \\
Mn & 5.52 & 1.93E-03 & 3.31E-07 & 1.15E+00 \\
P & 5.47 & 9.68E-04 & 2.95E-07 & 1.05E+00 \\
K & 5.12 & 5.46E-04 & 1.32E-07 & 1.10E+00 \\
\botrule 
\end{tabular}}
\begin{tabnote}
Abundances in CI meteorites of the twenty most abundant elements, their mass fractions in the chondritic mixture and their concentrations with respect to H. Values derived by scaling the values from\, \cite{lodders2010a} to the updated Si protosolar abundance  \cite{asplund2009,scott2015a,scott2015b}.
\end{tabnote}
\label{tab:meteoritic_abundances}
\end{table}

Alongside the stellar composition another important source of information on the elemental setup and the condensation processes of now-dispersed protoplanetary disks is supplied by the characterization of planetary materials, one of the most important of them being the meteorites. Meteorites are fragments of ancient planetary bodies that provide us a direct window into the processes that were occuring at the time planet were forming within the circumsolar disk.

As a first classification, we can divide meteorites into two main categories\cite{scott2007,nittler2016,wadhwa2020}: chondrites and achondrites. Chondrites are meteorites composed of materials that underwent no or limited modifications due to thermal or shock processes, their composition having been preserved mostly unaltered over the life of the Solar System (but see \cite{palme2014,palme2017} for a discussion of the differences between the various classes of chondrites). Achondrites are instead fragments of planetary bodies that underwent more marked or complete thermal processing and, as such, their composition has been significantly altered by the geophysical evolution of their parent bodies. 

The most extreme examples of such alterations are provided by the metallic achondrites and the stony achondrites, fragments of the cores and the surfaces, respectively, of planetary bodies that underwent geophysical differentiation like the Earth. This means that both these classes of meteorites lack part of the initial elemental budget of their parent bodies, e.g. the metals in the case of the stony achondrites. While achondrites can provide us important clues on the timescales and the sizes of the planetary bodies that were forming in the solar nebula\cite{scott2007,nittler2016,wadhwa2020}, for the purpose of understanding the initial elemental budget and condensation process of said planetary bodies our attention will focus on chondrites.

\begin{figure}[t]
\centering
\includegraphics[width=\textwidth]{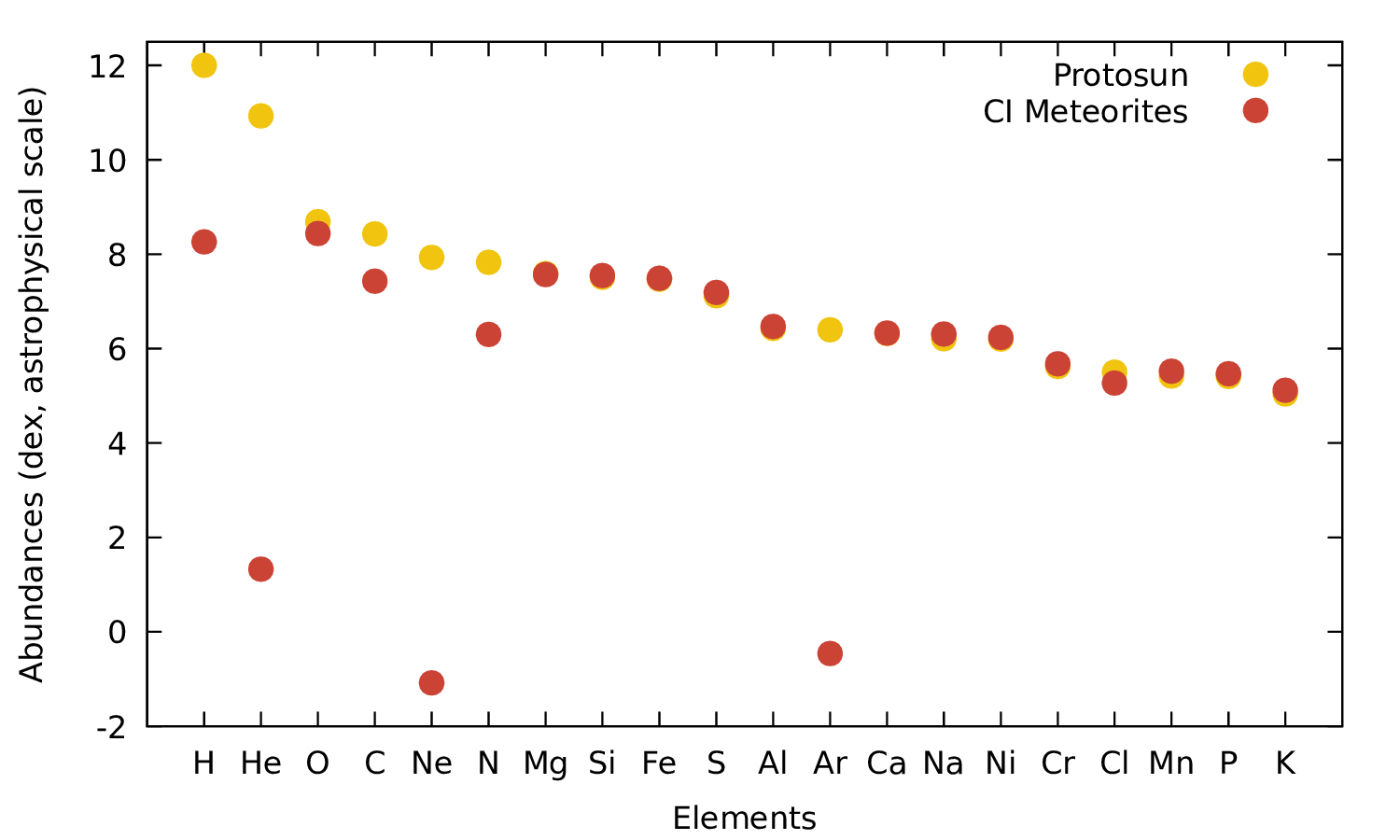}
\caption{Comparison between the protosolar elemental abundances (Table \ref{tab:protosolar_abundances},\,\cite{asplund2009,scott2015a,scott2015b}) and those of CI meteorites(Table \ref{tab:meteoritic_abundances},\,\cite{lodders2010a}) for the twenty cosmically most abundant elements.}\label{fig:abundance_comparison}
\end{figure}

Chondrites are divided into multiple classes\cite{palme2014,palme2017}, of which the CI chondrites are of particular importance. The elemental composition of CI chondrites is the one that most closely matches the solar composition except for the highly volatile elements C, O, N, H and the noble gases\footnote{The element Cl is notably underabundant in CI meteorites, see\,\cite{lodders2010a,palme2014,palme2017} for a discussion of the possible causes of this deficit.} (see Table \ref{tab:meteoritic_abundances} and Fig. \ref{fig:abundance_comparison}). All other classes of chondritic meteorites show decreased abundances of O and a  trend of comparative enrichment/depletion in more refractory/volatile elements with respect to CI meteorites \cite{palme2014,palme2017}, suggesting they formed at higher temperatures based on the condensation sequence of refractory elements (see Sect. \ref{sec:astrochemistry} and \cite{palme2014,palme2017} for a discussion). These trends are broadly consistent with similar trends observed in the presence of carbonaceous and volatiles elements in asteroids across the asteroid belt\cite{michtchenko2016}.

Meteoritic abundances are generally expressed in the cosmochemical abundance scale\cite{lodders2010a,palme2014,palme2017}: this scale works like the astrophysical abundance scale (see Sect. \ref{sec:host_stars}) in being a comparative abundance scale, the difference being that the reference abundance is provided by 10$^{6}$ atoms of Si (6 dex) instead of 10$^{12}$ atoms of H \cite{lodders2010a,palme2014,palme2017}. The two scales can be compared by equaling the abundance of Si atoms and scaling the abundances of all elements in the cosmochemical scale by the ratio between the abundance of Si in the astrophysical scale and 10$^{6}$ atoms of Si. Comparing the abundances of H in Tables \ref{tab:protosolar_abundances} and \ref{tab:meteoritic_abundances} immediately highlights the reason for such change in the reference element. 

The abundance of H atoms per atom of Si is almost four orders of magnitude lower in meteorites that in a protosolar mixture, as only a minimal fraction of the total budget of H in circumstellar disks in sequestered by refractory materials or, more generally, by solids. As discussed in Sect. \ref{sec:astrochemistry} and shown in Table \ref{tab:protosolar_abundances}, Si is a refractory element that is both abundant and condenses at high temperatures, making it a convenient reference for measuring abundances in solids. The comparison between meteoritic and protosolar values in Table \ref{tab:meteoritic_abundances} and Fig.  \ref{fig:abundance_comparison} provides several pieces of information. First, the abundances of all refractory elements in meteorities match the protosolar ones within 15 \% or 0.06 dex, the limit to the accuracy in the determination of meteoritic and stellar abundances (see\, \cite{lodders2010a,palme2014,palme2017} for a detailed discussion of the topic). 

This match is illustrated in Fig. \ref{fig:abundance_comparison} and indicates that, within the limits of the accuracy of the measures, in CI meteorites all elements belonging to the astrochemical refractory class are condensed in the same proportions as they were present in the solar nebula, which in turn indicates their complete condensation in the orbital region where the parent body of CI chondrites formed. Because of this match, meteoritic and Solar System study use the term \textit{chondritic composition} to indicate planetary material where the relative abundances of the refractory elements are in the same proportions as in the protosolar composition. As highlighted by Table \ref{tab:meteoritic_abundances} and Fig. \ref{fig:abundance_comparison}, a body with chondritic composition \textit{does not possess} protosolar abundances of the volatile elements H, C, O, N, and the noble gases, which are all underrepresented in meteorites.

Noble gases are present only in traces, with abundances in CI meteorites ten orders of magnitude lower than in the solar nebula. H, as mentioned before, in four orders of magnitude less abundant than in the solar nebula. The depletion of the elements O, C and N in meteorites is a function of their relative volatility: O, the comparatively more refractory of the three, is present in CI meteorites with about half the protosolar abundance, meaning that there is only half the total O budget available to form ices at lower temperatures. C and N are increasingly more volatile and are respectively present in CI meteorites with abundances that are about 9\% and 3\% their protosolar abundances.

Cometary data \cite{mumma2011,altwegg2019,oberg2020}, circumstellar disks \cite{pontoppidan2014,oberg2020}, and the interstellar medium \cite{gibb2004,oberg2011,eistrup2016,palme2017} jointly allow us to investigate the fate of the C, O and N not accounted for by meteorites. Before proceeding, an important point to keep in mind is that in general these sources of data do not provide absolute abundances but relative ones: the abundances of the different molecules are measured as relative abundances with respect to water, the most abundant volatile molecule. As a result, in order to convert these relative abundances to absolute ones the information on the abundance of water is required. This, in turn, makes it necessary to know the abundance of O available to form ices, i.e. not already sequestered in molecules involving refractory elements as illustrated by CI meteorites.

The information provided by the interstellar medium \cite{gibb2004,oberg2011,eistrup2016,palme2017} suggests that about 50-60\% of O available in volatile form should be in the form of H$_2$O. Since CI meteorites reveal that about half the total budget of O is sequestered in refractory form before the snow line of H$_2$O in a protosolar mixture, the O incorporated into H$_2$O should amount to about 25-30\% of the total protosolar O. When we scale by the abundance of H$_2$O, cometary data\cite{mumma2011,altwegg2019} reveals that about 1/3 of the protosolar budget of C is present in comets as ices, mainly CO and CO$_2$ as discussed in Sect. \ref{sec:astrochemistry}. While recent comparative observations support such a scenario of chemical inheritance of the volatile species in protoplanetary disks from the interstellar medium \cite{bianchi2019,drozdovskaya2019,oberg2020,turrini2021b}, it should be emphasized that also scenarios of chemical reset, where the volatile elements would recombine under different conditions of pressure and temperature with respect to the interstellar medium, are possible (see \cite{eistrup2016,oberg2020} for recent discussions)

When the abundance of C in ices is added to the C already present in CI meteorites, refractories and ices cumulatively account for about 40\% of the total budget of C. Recent observations of comet 67P Churyumov-Gerasimenko by the ESA mission Rosetta revealed that cometary dust is significantly richer in C than CI chondrites: specifically, the C/Si=6 and C/H$\approx$1 ratios measured in cometary dust suggest that about 60\% of protosolar C is present in comets as organic material \cite{bardyn2017,isnard2019,altwegg2019}. Such a high abundance of organic material in cometary dust means that organics represent, alongide rocks, metals and ices, one of the main solid phases within circumstellar disks. The investigation of prestellar and protostellar environments also confirms the rich chemistry of organic compounds (see \cite{oberg2020,turrini2021b} and references therein).

We can quantify the implications of such high abundance of organic material directly from the values reported in Table \ref{tab:meteoritic_abundances}: C accounts for 2.5$\times$10$^{-3}$ of the total mass in the protosolar mixture or, equivalently, about 18\% of the total mass of its Z elements (2.5$\times$10$^{-3}$/1.4$\times$10$^{-2}$). If 60\% of carbon is sequestered in organic compounds, this implies that organics represent about 10\% of the total budget of Z elements that can condense as solids in circumstellar disks (see also \cite{semenov2003} for similar considerations for the interstellar medium). This rough estimate does not account for the contributions of O, N and H to the mass fraction of organics in disks\cite{pollack1994}, yet cometary data suggests their mass contribution to be limited\cite{altwegg2019}.

Another information we can derive from cometary data and the intestellar medium is linked to N: as discussed before, only a minimal fraction ($\sim$3\%) of the protosolar budget of N is present in meteorites. Cometary ices appear to account for a commensurable fraction of N, likely no more than 10\% of its protosolar budget, mainly in the form of NH$_{3}$ ice. The investigation of the prestellar environments in molecular clouds shows how the majority of N is present as N$_2$ in protoplanetary disks, which condenses as ice at extremely low temperatures (see Sects. \ref{sec:astrochemistry} and \ref{sec:protoplanetary_disks}) and therefore remains in gas form across most of the disk extension \cite{pontoppidan2014,oberg2020}. 

Recently, the atmospheres of young A-type stars still surrounded by their protoplanetary disks \cite{kama2019} and the depletion patterns of elements in the gas of the disks \cite{mcclure2020} have been used to probe the composition of extrasolar solid materials, revealing that the condensation behaviour of elements around other stars follows the same general picture described by meteorites and comets in the Solar System. In particular, observations of S abundances in extrasolar materials in protoplanetary disks suggests that about 80-90\% of S in locked in refractory form\cite{kama2019}, in agreement with the estimated S abundance in cometary ices (10-20\% of the protosolar budget, depending on the abundance of water on which the abundances of the different ices are scaled\citep{mumma2011,turrini2021a}) and with the uncertainty on the determination of the abundance of S in meteorites.

Another source of information on extrasolar planetary materials comes from stars at the end of their life, specifically from the atmospheric contamination of white dwarf stars by the accretion of planetary bodies\cite{jura2014}. The study of polluted white dwarf stars reveals that the composition of the planetary bodies they accreted is overall consistent, in terms of abundances of both refractory elements and O, with the composition of meteorites and comets in the Solar System\cite{jura2014,doyle2019}, confirming how a large fraction of O is bounded with refractory elements and incorporated into rocky material as shown in the case of meteorites in Table \ref{tab:meteoritic_abundances}.

\section{Compositional structure of protoplanetary disks}\label{sec:compositional_structure}

By combining the information provided by the stellar composition, planetary materials and the thermal structure of protoplanetary disks as discussed in Sects. \ref{sec:protoplanetary_disks}, \ref{sec:host_stars}, and \ref{sec:planetary_materials} with the information on the condensation temperatures of the different molecules and materials discussed in Sect. \ref{sec:astrochemistry}, it becomes possible to quantitatively characterize the compositional structure of the planet-forming environments in protoplanetary disks.

\subsection{Metallicities of refractory materials, organics and ices}\label{sec:metallicities}

As discussed in Sect. \ref{sec:planetary_materials} and illustrated in Fig. \ref{fig:abundance_comparison}, the comparison of the elemental abundances in the protosolar mixture and CI meteorites shows that, with a 15\% accuracy\cite{lodders2010a}, all elements except for C, H, O, N and the noble gases are sequestered into solids before the snow line of H$_2$O. The sum of their mass fractions in the protosolar mixture reveals that refractory elements condensing in a CI chondritic mixture have Z$_{ref}$=$6.6\times10^{-3}$, i.e. about 47\% of the protosolar Z$_{proto}$=$1.41\times10^{-2}$. The comparison between the protosolar and chondritic mixtures highlights how, due to its high reactivity, O behaves both as a refractory and as a volatile element\cite{fegley2010,palme2014}. 

By subtraction of Z$_{ref}$ from Z$_{proto}$, we can see that volatile elements account for 53\% of Z$_{proto}$, i.e. Z$_{vol}$=$7.5\times10^{-3}$. About 18\% of Z$_{vol}$, however, is due to Ne that, as discussed in Sects. \ref{sec:astrochemistry}, \ref{sec:protoplanetary_disks} and \ref{sec:host_stars}, likely does not condense in the environment of protoplanetary disks\cite{fegley2010}. This means that the mass fraction of volatile elements that can condense in protoplanetary disks and contribute to the building blocks of planetary bodies amounts to Z$_{vol,cond}$=$6.2\times10^{-3}$. From these vaues it follows that the mass fraction of condensed material is globally lower than the Z value of the host star, e.g. lower than 1.4\% in the solar nebula, and the metallicity of the gas will globally remain higher than zero, i.e. we generally should not expect to have complete condensation of all Z elements in protoplanetary disks.

Another observation we can derive from the comparison of Z$_{ref}$ and Z$_{vol}$ is that, at full condensation of all Z elements (except Ne), their mass contributions to the planetary building blocks are about equal. As we discussed above, however, Z$_{vol,cond}$ is generally lower than Z$_{vol}$, meaning that the mass of planetary building blocks is generally dominated by refractory materials as rocks and metals. Furthermore, as discussed in Sect. \ref{sec:planetary_materials}, about 60\% of C appears locked by organics in protoplanetary disks, meaning that Z$_{org}$=$1.5\times10^{-3}$. By subtracting Z$_{org}$ from Z$_{vol,cond}$ we finally obtain Z$_{ice}$=$4.7\times10^{-3}$, i.e. the total mass fraction of condensable ices. The abundance of organics discussed here has been constrained considering only the contribution of C and using observational data from the Solar System. It should be noted, however, that values about twice as high of Z$_{org}$ were considered in previous studies of the interstellar medium\cite{semenov2003}. 

\begin{figure}[t]
\centering
\includegraphics[width=\textwidth]{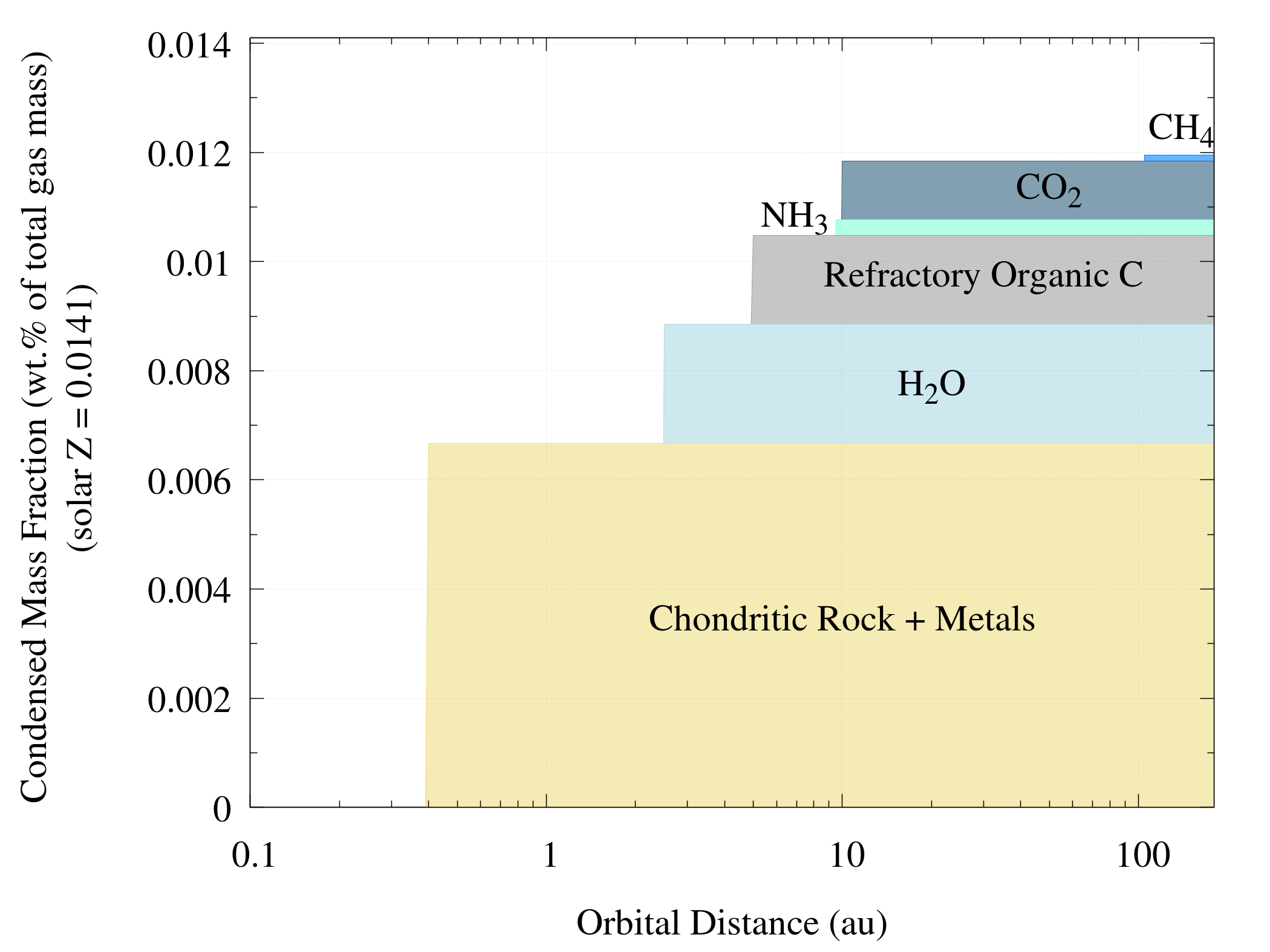}
\caption{Example of the compositional and solid mass fraction gradients in protoplanetary disks based on the disk model from \cite{turrini2021a} and discussed in Sects. \ref{sec:metallicities} and \ref{sec:compositional_model}.}\label{fig:compositional_gradient}
\end{figure}

The process of quantifying and balancing the mass contributions described here can be repeated for the different ices composing Z$_{ice}$ and the different refractory materials composing Z$_{ref}$ to derive the growth of the solid mass fraction in protoplanetary disks when crossing each condensation temperature or snowline\cite{semenov2003,oberg2011,madhusudhan2019,turrini2021a}, as shown in Fig. \ref{fig:compositional_gradient} and illustrated in the following example.

\subsection{Modelling the compositional structure of gas and solids}\label{sec:compositional_model}

The following illustrive example is derived from the protoplanetary disk model adopted in \cite{turrini2021a} and assumes $\approx$32\% of total O locked in water ([H$_2$O/H]=$1.7\times10^{-4}$), $\approx$6\% of O locked in CO ([CO/H]=$3.4\times10^{-5}$), $\approx$32\% of O locked in CO$_2$ ([CO$_2$/H]=$3.4\times10^{-5}$), $\approx$60\% of C locked in organics ([Organics/H]=$1.8\times10^{-4}$), $\approx$3.5\% of C locked in CH$_4$ ([CH$_4$/H]=$1.0\times10^{-5}$) based on the combined information from cometary data, the interstellar medium and astrochemical models\cite{eistrup2016}. 

In terms of N, this model assumes that about $33$\% of all N not sequestered by refractory materials  is in the form of NH$_3$ and the rest in the form of N$_2$ following the astrochemical models by \cite{eistrup2016}. Note that this partition of N between NH$_3$ and N$_2$ contains much more NH$_3$ than suggested by the interstellar medium, yet in the framework of this illustrative example this is not critical. The disk model adopted by \cite{turrini2021a} is characterized by the same temperature profile as the classical solar nebula, T = $T_0 \left( r/r_0\right)^{-\beta}$ where $T_0$=280 K and $\beta$=0.5, meaning this disk is warm by the standard of protoplanetary disks (see Sect. \ref{sec:protoplanetary_disks}). In such a warm disk, the ultra-volatile molecules CO and N$_2$ never condense and remain in gas form.

In the inner and hotter regions of the disk only refractory elements are condensed as rocks and metals (see Fig. \ref{fig:compositional_gradient}). It is worth reminding that, as discussed in Sect. \ref{sec:astrochemistry} and highlighted in the discussion on chondritic meteorites in Sect. \ref{sec:planetary_materials}, also refractory elements have their own condensation sequence\cite{lodders2003,fegley2010,palme2014,palme2017}. The bulk of the mass of refractory elements, however, is accounted by Fe, Si, Mg and S\cite{lodders2010a}. The most volatile of them, S, condenses at about 700 K (i.e. between 0.1 and 0.2 au for the adopted temperature profile) and by 1 au the disk is cold enough for all refractory elements to condense. As a consequence, we can assume that in the inner and hotter regions of the disk Z$_{solids}$=Z$_{ref}$=$6.6\times10^{-3}$ and the composition of solids is chondritic as a reasonable approximation. Studies focusing on the disk regions within 1 au, however, need to resolve the condensation sequence of refractory elements.

\begin{table}[t]
\centering
\tbl{Example of condensation sequence and $Z$ mass fractions of solids and gas in the solar nebula}
{\begin{tabular}{@{}cccc@{}} 
\toprule
Material & Snow line$^{\text a}$ &Z$_{solids}$ & Z$_{gas}$ \\ 
\colrule 
Rocks + Metals & 0.15 au & $6.6\times10^{-3}$  &  $7.5\times10^{-3}$  \\   
H$_{2}$O       & 2.5 au & $8.8\times10^{-3}$  &  $6.6\times10^{-3}$  \\
Org. C         & 5.0 au & $1.03\times10^{-2}$ &  $5.3\times10^{-3}$  \\
NH$_{3}$       & 9.5 au & $1.07\times10^{-2}$ &  $3.8\times10^{-3}$  \\
CO$_{2}$       & 10.5 au& $1.18\times10^{-2}$ &  $3.4\times10^{-3}$  \\
CH$_{4}$       & 105 au & $1.19\times10^{-2}$ &  $2.2\times10^{-3}$  \\
\colrule
Protosolar     &        &                     &  $1.41\times10^{-2}$  \\
\botrule 
\end{tabular}}
$^{\text a}$ The snow line of rocks and metals is arbitrarily located just outside the orbital distance where the disk temperature drops below the condensation temperature of S (see main text). 
\label{tab:z_values}
\end{table}

Moving outward, the first snow line we encounter is that of H$_2$O. As mentioned above, in this model water has an abundance relative to H of $1.7\times10^{-4}$, with a molecular weight of 18 atomic mass units. Given that in this relative abundance scale the abundance of H is 1 by definition and the atomic weight of H is 1, we can compute the mass fraction of H$_2$O with respect to H as $\left( 18 \cdot 1.7\times10^{-4} \right) / \left( 1 \cdot 1 \right) = 3.1\times10^{-3}$. As the disk gas is not composed only by H but includes also He and the Z elements, to obtain the mass fraction of water with respect to the total mass of the gas, we need to multiply the value just computed by the mass fraction of H in the protosolar mixture, i.e. $0.715$ (see Table \ref{tab:protosolar_abundances}). This gives us the mass fraction of water ice, which is $0.715 \cdot 3.1\times10^{-3}=2.2\times10^{-3}$. At the crossing of the H$_2$O snow line, therefore, Z$_{solids}$ grows to $8.8\times10^{-3}$.

Solid bodies and dust formed beyond the H$_2$O snow line will be composed of rocks, metals and water ice, as shown by Fig. \ref{fig:compositional_gradient}. The following snow line is the one of organics\cite{bergin2015,mordasini2016}, which in the model used in this example are assumed to be composed of C only. As the atomic weight of C is 12 and the abundance of organics with respect to H has been set to $1.8\times10^{-4}$, the mass fraction of organics with respect to the disk gas is $0.715 \cdot \left( 1.8\times10^{-4} \cdot 12 \right) / \left(1\cdot1\right)=1.5\times10^{-3}$. The process is repeated at each subsequent snowline to obtain the values reported in Table \ref{tab:z_values}, building the compositional structure shown in Fig. \ref{fig:compositional_gradient}. The metallicity of the gas Z$_{gas}$ in each compositional region is obtained subtracting Z$_{solids}$ from Z$_{proto}$.

\subsection{Effects of dust evolution and planetesimal formation}\label{sec:dust_and_planetesimals}

At the beginning of the life of protoplanetary disk the only solids populating them are dust grains. As a consequence, the solid mass fractions Z$_{solids}$ reported in Table \ref{tab:z_values} and shown in Fig. \ref{fig:compositional_gradient} will correspond to the dust-to-gas ratios in the different compositional regions of the disk. The dust suface density profile $\Sigma_{dust}$ can then be derived from the gas surface density profile $\Sigma_{gas}$ as:
\begin{eqnarray}
\Sigma_{dust}(r) & = & Z_{solids}(r) \cdot \Sigma_{gas}(r) \nonumber \\ & = & Z_{solids}(r) \cdot \Sigma_{c} \left( r/r_{c})\right)^{-\gamma} exp \left[ -\left( r/r_{c} \right)^{2-\gamma}\right]\label{eqn:solids_density}
\end{eqnarray}
where, as discussed in Sect. \ref{sec:protoplanetary_disks}, r$_c$ is the characteristic radius of the disk and $\Sigma_{c}$ the gas surface density at the characteristic radius.

The initial dust density distribution described by Eq. \ref{eqn:solids_density} matches the distribution of solids in protoplanetary disks only at the very beginning of their life. As introduced in Sect. \ref{sec:planet_formation}, dust orbits the central star on keplerian orbits while the gas is partially supported by the pressure gradient within the protoplanetary disk\cite{weidenschilling1977} and, as a result, it orbits with sub-keplerian velocity. This differential orbital motion between gas and dust, with the latter moving faster than the former, results in the dust experiencing an aerodynamic drag. Small dust grains (roughly micron-to-mm sized) will experience the strongest drag effect and will quickly dynamically thermalize, become comoving with the gas.

Larger grains (roughly mm-to-cm sized) will experience the headwind of the gas and will lose angular momentum, starting to radially drift inward with respect to the gas. As introducted in Sect. \ref{sec:planet_formation}, the size-dependent nature of the drag process results in the differential motion between dust grains of different sizes. This differential motion plays an important role in enhancing both the dust growth and the planetesimal formation processes (see \cite{johansen2014,johansen2017} for detailed discussions). From the perspective of the dust-to-gas ratio, this means that the distribution of the dust with respect to the gas will not necessarily remain as described by Eqs. \ref{eqn:dust_density} and \ref{eqn:solids_density}, with several important implications for the compositional structure of the protoplanetary disk.

First, the dust distribution of protoplanetary disks will evolve over time as the dust drifts inward toward the inner regions while the outer regions become dust-depleted\cite{testi2014,dipierro2016,toci2021}. This process is observationally confirmed by the comparison of the radial extensions of gas and dust in protoplanetary disks, which reveals how the dust is two to four times more compact than the gas\cite{isella2016,ansdell2018,facchini2019}. Since the dust is expected to grow and, due to the size-dependent nature of gas drag, slow down\cite{weidenschilling1977} and pile up in the inner regions of disks, this process will globally result in enhanced mass fractions of solids (i.e. the solid metallicity Z$_{solids}$) up to the outer radius of the dus distributiont\cite{mordasini2009,turrini2021a}. Because of the differential motion of dust grains with different sizes, however, this enhancement will not necessarily be homogeneous throught the disk (see \cite{isella2016} for an illustrative example).

Second, the drifting dust will deliver both refractory and volatile elements inward with respect to the different snow lines at which they condense as refractory materials, organics and ices. As a consequence, part of the condensed mass will be released back to the gas phase, increasing its metallicity Z$_{gas}$. It should be emphasized how this enrichment will differ across the extension of the disk, as it will selectively concern only the sublimating element: as an example, inside the CO snow line Z$_{gas}$ will be enhanced do to the release of C and O in the gas, while inside the N$_2$ snowline the enhancement will be due to N only.  While the composition of solids throughout the disk will remain the same over time, that of gas will evolve. Planets accreting gas while crossing the metallicity-enhanced regions will possess higher metallicity values and will be selectively enriched in specific elements with respect to planets accreting gas not influenced by this process\cite{bosman2019,boothr2019,schneider2021a,schneider2021b}.

It should be noted that the metallicity enhancement process caused by the dust drift and evaporation can be countered by the thermal evolution of the disks, which will radiatively cool down during their lifetimes and experience an inward drift of their snow lines\cite{panic2017,eistrup2018}. As the enriched regions are crossed by the migrating snow lines, the sublimated elements will condense once again on the grains and revert the metallicity of the gas Z$_{gas}$ to its original state. The extent to which the two processes cancel out will depend on the balance between the dust and snow lines timescales of inward drift. 

It is important to emphasize, however, that the above mentioned processes are relevant only as long as most of the mass of solids throughout protoplanetary disks is in the form of dust grains. Since the dust is concentrated and converted into planetesimals during its drifting motion across the gas\cite{chambers2010b,johansen2014,johansen2017}, over time more and more solid mass will be converted into planetesimals, decresing the solid mass fraction accounted for by dust. As planetesimals are characterized by less favourable surface-to-volume ratios with respect to dust, when crossing a snow line only a minimal fraction of their mass will be directly affected by the higher temperature and sublimate or provide a site for the cooled volatiles to condense on \cite{turrini2021a,turrini2021b}.

Planetesimal formation, therefore, will act to lower the efficiency ot the gas enrichment process as well as the effects of the drifting snow lines described above. Once again, the net effect on the disk will depend on the balance between the timescales of the processes described above and that of the conversion of the dust into planetesimals. The latter process is expect to occur over a timescale of 1 Myr or less based on the most recent theoretical frameworks\cite{chambers2010b,johansen2014,johansen2017}, a timescale consistent with that suggested by observational data in the Solar System and protoplanetary disks. 

Specifically, the planetesimal formation timescales derived from radiometric dating of meteorites\cite{scott2007,nittler2016,wadhwa2020} are of the order of a few 10$^{5}$ years for the first generations of planetesimals to appear. In parallel, the observations of protoplanetary disks consistently show that disks older than about 1 Myr have dust masses too low to explain the formation of the large number of multi-planet systems we know to date \cite{manara2018}, suggesting the conversion of large amounts of dust into planetesimals. Overall, these observational constraints suggest the possibility that the compositional structure of protoplanetary disks can stop evolving within the first 1 Myr, i.e. as soon as most of the initial mass of dust is converted into planetesimals\cite{turrini2021a,turrini2021b}.

\section{Compositional signatures of planet formation}

The template protoplanetary disk we built as our working example in Fig. \ref{fig:compositional_gradient} illustrates how the solids and, by subtraction from the stellar composition, the gas at different distances from the star will be characterized by different compositions and elemental budgets: a body formed at 8 au will include in its composition chondritic rocks and metals, water and organics, and will be richer in C and O with respect to a body formed at 2 au, which will include only the O and C present in chondritic material (see Table \ref{tab:meteoritic_abundances}).

\begin{figure}[ht]
\centering
\includegraphics[width=0.70\textwidth]{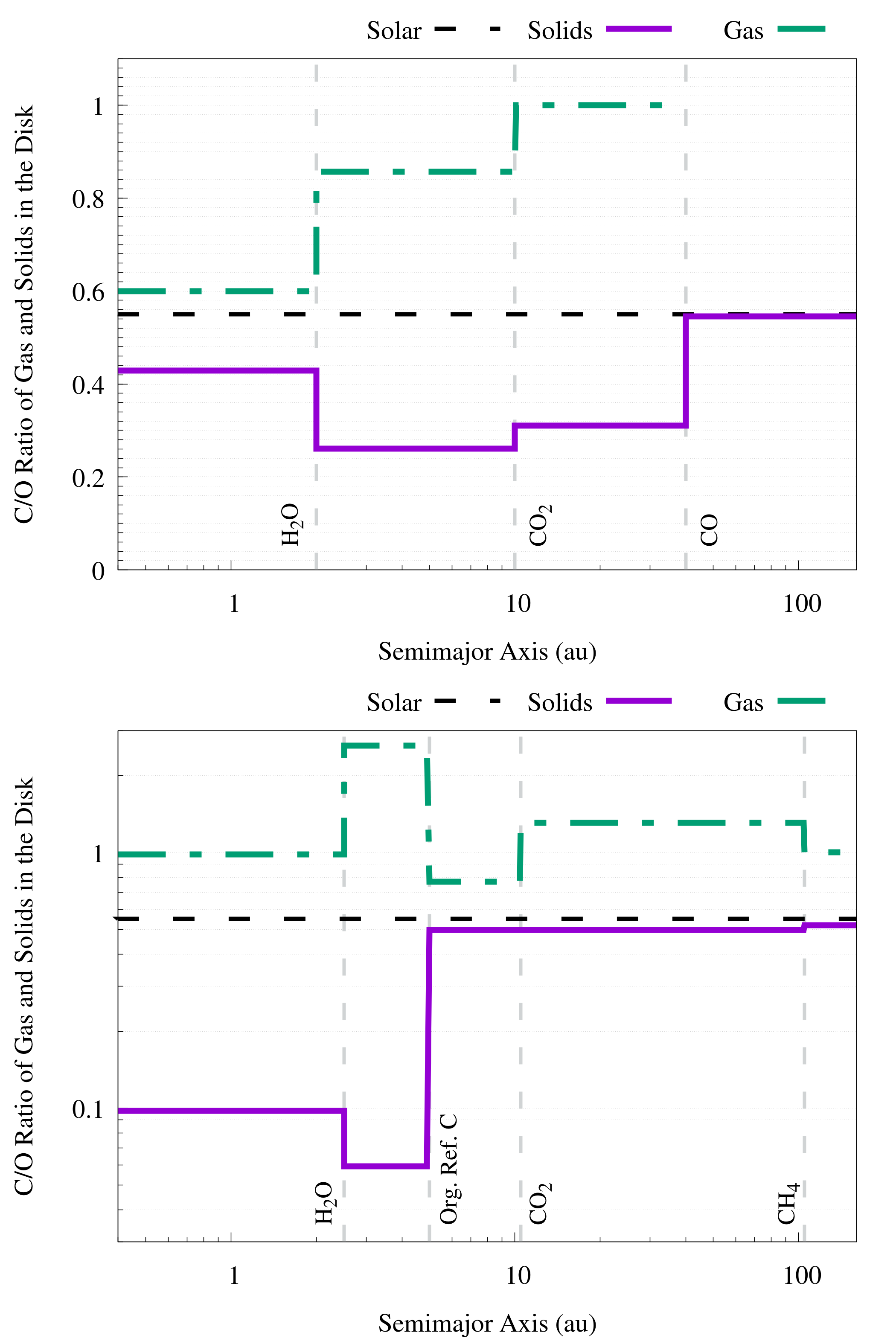} 
\caption{Illustrative examples of the C/O ratios of gas and solids across protoplanetary disks. The top panel shows the original distribution of C and O considered by\, \cite{oberg2011} when discussing the use of the planetary C/O ratio as a window into the formation and migration histories of planets. The bottom panel shows the updated C/O ratios of gas and solids derived by\,\cite{turrini2021a} and accounting for the joint information provided by Solar System, interstellar medium and polluted white dwarfs. Note that the bottom plot is based on a warmer disk, which is why it does not include the CO snow line.}\label{fig:c_o_ratio}
\end{figure}

These differences can be expressed in terms of the abundances ratios among the different elements, among which O and C were the first used to characterize the composition of solids and gas in protoplanetary disks\cite{oberg2011}. Since then, a rich literature focusing on the study of the C/O abundance ratio in protoplanetary disks and exoplanets has developed (see \cite{madhusudhan2016,madhusudhan2019,oberg2020,turrini2021b} for recent reviews on the subject). The basic idea is illustrated in Fig. \ref{fig:c_o_ratio}, where the upper plot reproduces the original figure from \cite{oberg2011} while the bottom one performs the same computation for the template disk of Fig. \ref{fig:compositional_gradient}. 

As the upper plot of Fig. \ref{fig:c_o_ratio} illustrates, at the crossing of the snow lines of their various molecules the abundance ratio between C and O will vary in both the gas and solids. As an example, when crossing the H$_2$O snow line solids will be enriched in O while gas will be depleted. As the H$_2$O snow line does not affect C, the C/O ratio of solids will decrease (constant C, increased O) while that of gas will increase (constant C, decreased O).  At the crossing of the CO snow line, the most volatile carrier of C and O, the C/O ratio of solids will match the stellar one (assumed by \cite{oberg2011} to be equal to the solar one of 0.55) as all C and O will be in solid form. Before the CO snow line, the C/O ratio of the gas will be 1 as CO molecules contain equal quantities of the two elements (see Fig. \ref{fig:c_o_ratio}, upper plot). 

As a result of the higher relative volatility of C with respect to O, the C/O ratio of solids will be substellar (O condenses in solids earlier than C) while that of the gas will be superstellar. In principle, therefore, one can use the C/O ratio of a planetary body to constrain its formation region: gas-dominated giant planets should be characterized by superstellar C/O ratios. Because of the monotonic trend of the C/O ratio of the gas, the greater the C/O ratio of the giant planet the farther away from the star it should have formed (see Fig. \ref{fig:c_o_ratio}, upper plot). Terrestrial planets and super-Earths should be instead characterized by substellar C/O ratio unless they formed beyond the CO snow line. While the basic idea is sound, the interpretation of the C/O ratio is not necessarily as straightforward as it appears.

\subsection{The limits of the C/O ratio}

For the interpretation of the C/O ratio to be unequivocal, in fact, the compositional structure of protoplanetary disks should satisfy the following two implicit conditions: a)  the compositional signatures created by the different snow lines should not be degenerate (i.e. multiple regions should not produce the same signatures), and b) the differences in the C/O ratios of gas and solids across the different compositional regions should be large enough to be measurable. As the bottom panel of Fig.  \ref{fig:c_o_ratio} illustrates, these two conditions are not necessarily satisfied in disks.  Furthermore, planetary bodies would need to satisfy two additional conditions: c) they should only accrete local material while forming, where by local material we refer to material from the same compositional region of the disk, and d) they should migrate only after they reached almost their final mass.

When we compute the C/O ratio of the gas and solids in the template protoplanetary disk from Fig. \ref{fig:compositional_gradient}, accounting for the compositional information provided by all astrophysical sources discussed in Sect. \ref{sec:compositional_structure}, we can  see that there is no monotonic trend in the C/O ratio of the gas, which instead oscillates around a value of 1 (see Fig. \ref{fig:c_o_ratio}, bottom panel; also in this case the stellar C/O ratio is assumed to match the solar one). When looking at the solids we can see that, as soon as organics and water are added to their composition, their C/O ratio jumps to an almost stellar value (see Fig. \ref{fig:c_o_ratio}, bottom panel). Since the snow lines of organics and water are the closest to the star\cite{bergin2015}, the C/O ratio of solids across most of the extension of the protoplanetary disk will be indistinguishable from the stellar one.

Even in protoplanetary disks satisfying the conditions a) and b) discussed above, however, the information encoded in the C/O ratio is not necessarily unequivocal to interpret due to the process of planetary migration. As introduced in Sect. \ref{sec:planet_formation}, planetary bodies embedded in protoplanetary disks will migrated due to their interactions with the gas. Planetesimals will experience an inward drift due to the aerodynamic drag of the gas whose intensity will vary with the size of the planetesimals\cite{weidenschilling1977}. Small planetesimals can diffuse inward with respect to their original compositional region and be accreted by other planetary bodies along their path or by terrestrial planets during their final assembly long after the dispersal of the disk\cite{chambers2010b}, altering their C/O ratio with respect to that of the orbital region where they formed.

Large terrestrial planets formed while still embedded in the circumstellar disk will experience what is known as Type I migration\cite{chambers2009,nelson2018} and are expected to migrate over significant orbital fractions of the disk extension\cite{mordasini2015,johansen2019}. Even in disk characterized by limited or no inward diffusion of the planetesimals, therefore, growing planets with masses encompassing the range of super-Earths will cross different compositional regions before reaching their final orbital location and can accrete solid material that can alter their C/O value and mask their original formation region. 

As an example, when we consider the disk depicted in the upper plot of Fig. \ref{fig:c_o_ratio}, we can see that a super-Earth planet accreting half of its mass beyond the CO snow line and the other half between the CO$_2$ and CO snow lines can end up with a C/O ratio resembling that characteristic of planets formed within the H$_2$O snow line (see Fig. \ref{fig:c_o_ratio} and \cite{turrini2015,turrini2018}). In principle, such ambiguity can be addressed by combining the information provided by the C/O ratio with the one arising from the planetary density, as a body formed beyond the CO$_2$ snowline will be richer in ices than a body formed within the H$_2$O snowline. As the near-identical densities the asteroid Ceres and the dwarf planet Pluto reveal, however, also the information supplied by density can be degenerate (see \cite{turrini2018} for a discussion).

The situation is similar for giant planets, which will experience both Type I migration during the growth of their core and Type II migration during the runaway accretion of their massive gaseous envelopes\cite{chambers2009,dangelo2010,nelson2018,johansen2019,tanaka2020}. During the runaway gas accretion phase, giant planets can accrete large masses of planetesimals or undergo giant impacts with other planetary bodies, acquiring in the process  markedly supersolar metallicities\cite{shibata2020,turrini2021a,ogihara2021} consistent with those estimated for extrasolar giant planets \cite{thorngren2016} and observed in the giant planets of the Solar System \cite{atreya2018}. 

Extracting details on the migration history of giant planets from their metallicity, however, is not straightforward, as the correlation between the extent of the migration and the metallicity is degenerate with respect to the migration rate\cite{shibata2020}. As a result, a rapid migration over a shorter distance can produce the same metallicity of a slower migration occurring over a longer distance\cite{shibata2020}. When the metallicity is decomposed into the abundances of the different elements and the C/O ratio is computed, moreover, its values can possess a limited diagnostic power, particularly for giant planets that experienced extensive migration, and only allow to discriminated low-metallicity, gas-dominated giant planets from high-metallicity, solid-enriched giant planets\cite{turrini2021a,turrini2021b} without providing information on the migration and formation histories (see Fig. \ref{fig:elemental_ratios}). 

\subsection{Expanding the inventory of elemental ratios}

Following the increasing number of elements and molecules that are being identified and whose abundances are starting to be estimated in exoplanetary atmospheres\cite{tsiaras2018,pinhas2019,welbanks2019,giacobbe2021}, recent studies started exploring the information delivered by elemental ratios involving other elements than C and O. The initial focus of these studies has been on the elements N (see   \cite{oberg2019,bosman2019,cridland2019,turrini2021a}) and S (see \cite{turrini2021a}),  cosmically abundant elements at the opposite ends of the volatility spectrum. As discussed in Sects. \ref{sec:astrochemistry}, \ref{sec:planetary_materials} and \ref{sec:compositional_model}, N is a highly volatile element whose bulk mass remains in gas phase for most of the extension of protoplanetary disks. S is instead a refractory element whose bulk mass condenses in solid form quite close to the star.

\begin{figure}[t]
\centering
\includegraphics[width=0.75\textwidth]{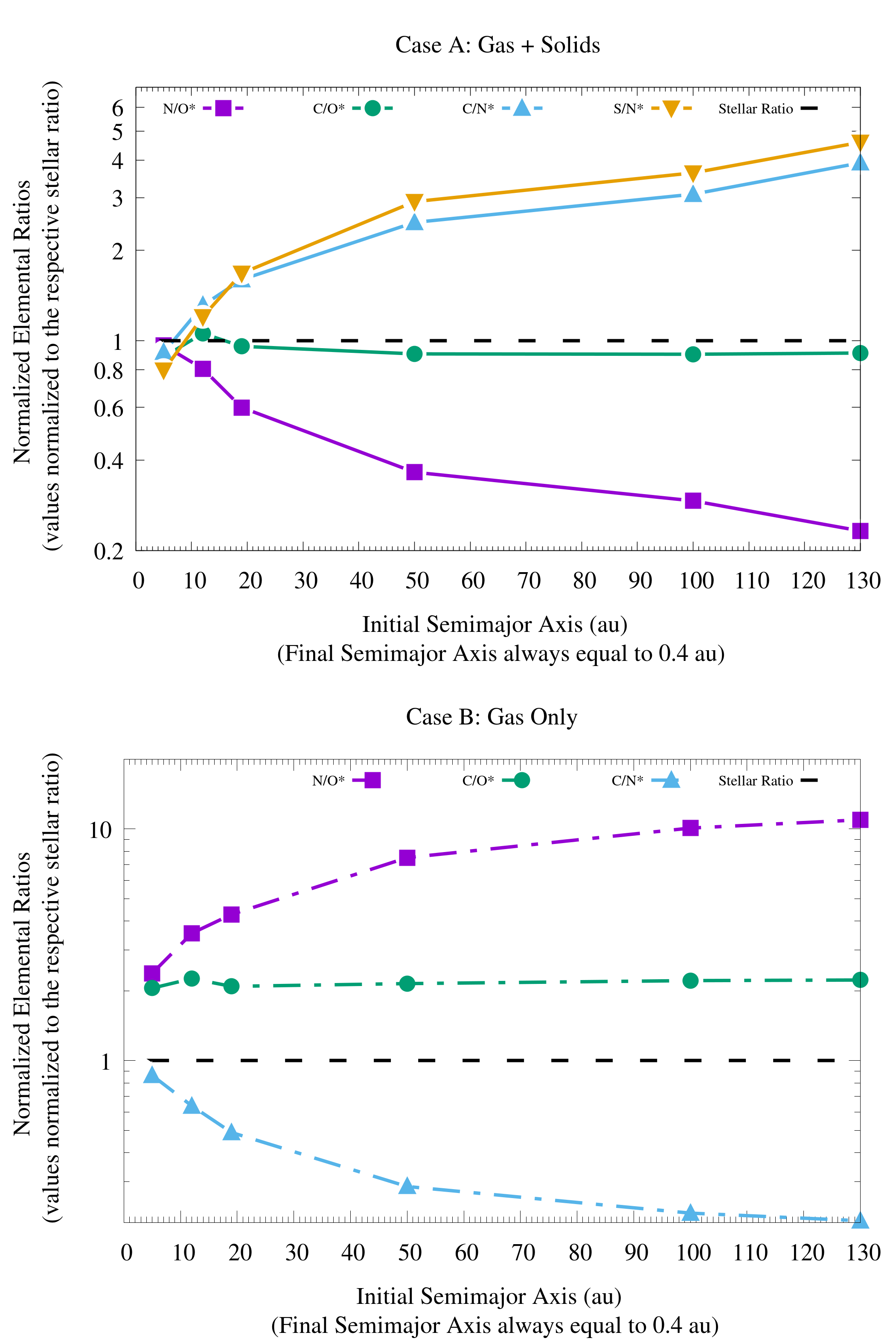} 
\caption{Elemental ratios of giant planets migrating from different starting positions to become hot Jupiters while accreting both gas and planetesimals (top plot) or only gas (bottom plot) (data from the simulations of\, \cite{turrini2021a}). The elemental ratios are normalized to the relevant stellar ones so a ratio of 1 means that it is equal to the stellar ratio.}\label{fig:elemental_ratios}
\end{figure}

Fig. \ref{fig:elemental_ratios} shows a comparison of the elemental ratios C/O, N/O, C/N, and S/N computed for a set of giant planets starting their formation at orbital distances compatible with the observations of young planets embedded into protoplanetary disks \cite{alma2015,isella2016,fedele2017,fedele2018,andrews2018,long2018} and migrating to become hot Jupiters \cite{turrini2021a}. The top plot of Fig. \ref{fig:elemental_ratios} shows the elemental ratios of the giant planets when they accrete both planetesimals and gas during their migration. From left to right, the migration scenarios are associate to final planetary metallicity values going from the stellar one to eight times that of the star. The bottom plot shows the elemental ratios of giant planets accreting only gas and characterized by a sub-stellar metallicity. The compositional structure of the protoplanetary disk is similar to that discussed in Sect. \ref{sec:compositional_model} but assumes a more realistic 9:2 partition between N$_2$ and NH$_3$\cite{turrini2021a}.


As can be immediately seen, the C/O ratio in both plots shows a flat profile with almost no sensitivity to how far the giant planets start their migration and to their actual metallicity. On the contrary, thanks to the higher contrast in volatility of N with respect to C and O, the C/N and N/O ratios show monotonic trends that deviate the more from the stellar values the farther away the giant planets start their migration. Since solids are richer in C and O than in N (see Sect. \ref{sec:compositional_model} for a discussion), the C/N ratio grows and the N/O ratio decreases with migration for giant planets where planetesimals contribute the most to the metallicity\cite{turrini2021a,turrini2021b}. 

For giant planets dominated by gas accretion the C/N ratio decreases while the N/O ratio increases with the extent of migration\cite{turrini2021a,turrini2021b}. It should be noted that the planetary elemental ratios in Fig. \ref{fig:elemental_ratios} are normalized to the relevant stellar elemental ratios\footnote{As discussed in Sect. \ref{sec:host_stars} (see also \cite{turrini2021b} and references therein), the various planet-hosting stars do not necessarily have elemental abundances matching the solar ones. As a consequence, the abundance ratios can provide  meaningful information only when compared to those of the host star, not with the values characteristic of the Sun.}. This normalization allows to plot the different ratios on a common scale and highlights how, in this normalized scale, solid-enriched giant planets will possess C/N* $>$ C/O* $>$ N/O*\cite{turrini2021a,turrini2021b}. Giant planets dominated by gas accretion will instead be characterized by N/O* $>$ C/O $>$ C/N*.\cite{turrini2021a,turrini2021b}. The S/N elemental ratio follows a monotonic trend similar to that of C/N (see Fig. \ref{fig:elemental_ratios}), but the more refractory nature of S with respect to C allows to extract additional information from the comparison of the two ratios.

Specifically, S/N* will be greater than C/N* when the metallicity is mainly due to the accretion of solids, as solids will contain more S than C (see Sect. \ref{sec:compositional_model}). When the accretion of gas significantly contributes to the planetary metallicity, the C/N* ratio will be greater than the S/N* ratio due to the higher volatility of C (see Fig. \ref{fig:elemental_ratios}). Therefore, the comparison of the normalized S/N* and C/N* ratios can allow to discriminate giant planets that accreted gas whose metallicity was enhanced by the drift and evaporation of ice-rich dust and pebbles across their respective snow lines (see Sect. \ref{sec:dust_and_planetesimals}) from those that were not affected by this process\cite{turrini2021a,turrini2021b}.

Finally, because of the high contrast in volatility between S and N, the S/N ratio is directly proportional to the contribution of planetesimals and solids to the planetary metallicity and, as such, can be used as a proxy of the latter\cite{turrini2021a,turrini2021b}. Before concluding, it is worth noting that the properties discussed for the S/N ratio are not exclusive of S but are common to all refractory elements. This fact, combined with the use of normalized ratios, can simplify the comparison between planets orbiting different stars since different ratios (e.g. Ca/N*, Ti/N*, Na/N*, Si/N*, Fe/N*) can be used to derive the same information, meaning that the choice of the specific elemental ratio to be adopted can be tuned to the observational data available on the planets and their stars.

In conclusion, by normalizing the  elemental ratios of giant planets to those of their stars and by comparing the behaviours of ratios of elements characterized by different volatility, it is possible to more robustly constrain the orbital migration and to determine the main source of the planetary metallicity, gas or solids, even without accurated knowledge of the planetary mass and radius (and, consequently, of the metallicity itself) or of the full planetary elemental budget.


\section{Future outlooks and concluding remarks}

The number of topics we touched upon during this brief dive into the compositional dimension of planet formation highlights how many different yet complementary pieces of information concur in producing our understanding of the composition of planets and how it comes to be. Rather than reviewing the most up to date observational data and theoretical models, the focus of our discussion has been on the process through which these different pieces of information can be connected and integrated into a unified picture. This approach was motivated by two reasons.

The first one is that, as mentioned at the beginning of this chapter, discussing in detail each of these interconnected topics would be impossible in the framework of this single chapter. The second one is that all discussed topics are evolving at an extremely rapid pace thanks to the increasingly detailed data provided by existing observational facilities both on ground and in space. The incoming launch of the James Webb Space Telescope \cite{cowan2015,greene2016} and, in a few years, that of the ESA space mission Ariel \cite{tinetti2018,turrini2018,tinetti2021} promise an unprecedented level of detail in the characterization of exoplanets. In parallel, in the incoming decade the Square Kilometre Array (SKA) promises as large a revolution in the compositional study of protoplanetary disks\cite{lazio2004,testi2015,codella2015} as that currently taking place thanks to the Atacama Large Millimetre/sub-millimetre Array (ALMA).

While it is difficult, if not impossible, to anticipate the challenges that the new observational data and discoveries will pose in the coming years, this chapter aims at providing the readers with the conceptual tools needed to confront them. I hope that the readers will find these conceptual tools both useful and effective in supporting their efforts to investigate the relationship between the composition of planetary bodies and that of their formation environment.



\bibliographystyle{ws-rv-van}
\bibliography{Turrini-Bibliography}

\end{document}